\documentclass[a4paper,12pt]{article}

\usepackage{amsmath}
\usepackage{amssymb}
\usepackage[dvips]{graphicx}
\usepackage[dvips]{psfrag}
\usepackage{bm}

\newcommand{\id}{{\bf 1}} 
\newcommand {\be}{\begin{equation}}
\newcommand {\ee}{\end{equation}}
\newcommand {\bea}{\begin{eqnarray}}
\newcommand {\eea}{\end{eqnarray}}
\newcommand {\nn}{\nonumber}
\newcommand {\tr}{{\rm tr}}

\newcommand{\cO}{{\cal O}}

\newcommand{\ket}{\rangle}
\newcommand{\bra}{\langle}
\newcommand{\vev}[1]{\left\langle #1 \right\rangle}

\begin{document}
\thispagestyle{empty} \addtocounter{page}{-1}
\begin{flushright}
OIQP-12-08\\
RUP-12-8
%
\end{flushright} 
\vspace*{1cm}

\begin{center}
{\large \bf  New critical behavior in a supersymmetric \\double-well matrix model}\\
\vspace*{2cm}
Tsunehide Kuroki$^*$ and Fumihiko Sugino$^\dagger$\\
\vskip0.7cm
{}$^*${\it Department of Physics, Rikkyo University, }\\
\vspace*{1mm}
{\it Nishi-Ikebukuro, Tokyo 171-8501, Japan}\\
\vspace*{0.2cm}
{\tt tkuroki@rikkyo.ac.jp}\\
\vskip0.4cm
{}$^\dagger${\it Okayama Institute for Quantum Physics, } \\
\vspace*{1mm}
{\it Kyoyama 1-9-1, Kita-ku, Okayama 700-0015, Japan}\\
\vspace*{0.2cm}
{\tt fumihiko\_sugino@pref.okayama.lg.jp}\\
\end{center}
\vskip2cm
\centerline{\bf Abstract}
\vspace*{0.3cm}
{\small 
We compute various correlation functions at the planar level in a simple supersymmetric matrix model, 
whose scalar potential is in shape of a double-well. 
The model has infinitely degenerate vacua parametrized by 
filling fractions $\nu_\pm$ representing the numbers of matrix eigenvalues around the two minima of the 
double-well. 
The computation is done for general filling fractions corresponding to 
general two-cut solutions for the eigenvalue distribution.  
The model is mapped to the $O(n)$ model on a random surface with $n=-2$, and 
some sector of the model is described by two-dimensional quantum gravity with $c=-2$ matter or 
(2,1) minimal string theory. 
For the other sector in which such description is not possible, we find new critical behavior of 
powers of logarithm for correlation functions.    
We regard the matrix model as a supersymmetric analog of the Penner model, and 
discuss correspondence of the matrix model to two-dimensional type IIA superstring theory 
from the viewpoint of symmetry and spectrum. 
In particular, single-trace operators in the matrix model 
are naturally interpreted as vertex operators in the type IIA theory. 
Also, the result of the correlation functions implies that the corresponding type IIA theory has a 
nontrivial Ramond-Ramond background.   
}
\vspace*{1.1cm}



\newpage

\section{Introduction}
Matrix models given by dimensional reduction of supersymmetric Yang-Mills theories have been attractive 
frameworks capturing nonperturbative aspects of superstring/M theory, 
which are based on conjectures~\cite{Banks:1996vh,Ishibashi:1996xs,Dijkgraaf:1997vv} and on AdS/CFT 
correspondence~\cite{Maldacena:1997re,Itzhaki:1998dd}.  
Although such matrix models have been vigorously investigated since these proposals, 
it is still hard to unveil nonperturbative string vacua and understand their essential properties. 
  
In this situation, we consider a simple supersymmetric matrix model discussed in~\cite{Kuroki:2009yg}, 
which is a zero-dimensional matrix model analogous to two-dimensional Witten's model containing 
a real scalar field and a Majorana fermion~\cite{Witten:1982df}. 
Its scalar potential takes the form of a double-well. 
This model seems much more tractable compared to the matrix models of supersymmetric Yang-Mills type, and 
we hope the analysis here is helpful to get insights into nonperturbative dynamics of strings  
not ever clarified and   
in particular crucially related to target-space supersymmetry. 
Ref.~\cite{Kuroki:2009yg} found 
that the model has infinitely degenerate supersymmetric vacua at the planar level. The degeneracy is 
parametrized by filling fraction $\nu_+$ ($\nu_-$) corresponding to 
the number of eigenvalues of a scalar matrix $\phi$ around the right minimum (the left minimum) of the 
double-well. The support of the eigenvalue distribution is two intervals for general $\nu_\pm$ (two-cut 
solutions). 

In this paper, we compute various correlation functions of the matrix model, 
and investigate their critical behavior as the two intervals of the eigenvalue distribution 
approach to touch each other.   
Although this model is equivalent to a matrix model for the $O(n)$ model on a random surface 
with $n=-2$~\cite{David:1985et,Kostov:1987kt,Kostov:1988fy,Kostov:1992pn}, 
analysis ever done in the literature is for one-cut solutions~\footnote{
For example, see refs.~\cite{Eynard:1995nv,Eynard:1995zv,Borot:2009ia}.
}, where the eigenvalue distribution 
has the support of a single interval. 
Thus, the analysis here would be also helpful to consider a multi-cut case in general $O(n)$ matrix models. 
As a result of the analysis, we find critical behavior of powers of logarithm 
in planar correlation functions 
of odd powers of the scalar matrix and of fermionic matrices. 
Such behavior has not ever been seen 
in the literature, although it is somewhat reminiscent of the logarithmic scaling violation in 
a matrix quantum mechanics called the 
$c=1$ matrix model~\cite{Kazakov:1988ch,Brezin:1989ss,Gross:1990ay,Ginsparg:1990as}~\footnote{For a review, 
see ref.~\cite{Klebanov:1991qa}.
}. 
    
The Penner model~\cite{Distler:1990mt} 
and its extension including source terms for ``tachyons'' 
(the Kontsevich-Penner model)~\cite{Imbimbo:1995yv,Mukhi:2003sz}    
are zero-dimensional matrix models whose critical behavior is the same as 
the $c=1$ matrix model~\footnote{
Also is the normal matrix model~\cite{Alexandrov:2003qk}, which corresponds to $c=1$ noncritical 
strings on $S^1$ with a general radius.
}. 
They describe bosonic string theory in two-dimensional target space: 
$(\mbox{Liouville direction})\times (\mbox{$S^1$ with the self-dual radius})$. 
It suggests us an intriguing direction to regard our supersymmetric matrix model 
as a supersymmetric version 
of the Penner model and 
to consider correspondence of the matrix model to two-dimensional string theory 
with target-space supersymmetry. 
Indeed, two-dimensional type II superstring theory with the same target space, 
(Liouville) $\times$ ($S^1$ with the self-dual radius), 
is constructed~\cite{Kutasov:1990ua,Murthy:2003es,Ita:2005ne,Grassi:2005kc}, 
where the target-space supersymmetry can exist only at the self-dual radius of the circle.    

The rest of this paper is organized as follows. 
After a short review of the supersymmetric double-well matrix model in the next section, 
we compute the one-point, two-point and three-point correlation functions among operators of the scalar 
matrix $\frac{1}{N}\tr\,\phi^n$ at the planar level, 
and investigate their critical behavior 
in sections~\ref{sec:1pt_phi}, \ref{sec:2pt_phi} and \ref{sec:3pt_phi}, respectively. 
The two-point functions of fermionic matrices are calculated in section~\ref{sec:2pt_psi}.    
We summarize the results obtained so far, and discuss correspondence of the matrix model 
and two-dimensional type IIA superstring theory from the viewpoint of symmetry and spectrum 
in section~\ref{sec:discussion}. 
Analysis of Schwinger-Dyson equations for the matrix model is presented in appendix~\ref{app:SD}. 
We discuss dependence on the filling fractions 
of correlation functions among operators of the scalar 
matrix $\phi$ in appendix~\ref{app:ff}.  
Some useful formulas for the calculation is given in appendix~\ref{app:formulas}.
 
The next paper~\cite{kuroki-sugino} is devoted to computing various amplitudes in the type IIA theory, and 
to discuss the correspondence in detail 
by comparing the amplitudes in both sides of the matrix model and the type IIA 
theory.

\section{Supersymmetric double-well matrix model} 
\label{sec:dbmm}
\setcounter{equation}{0}
Let us start with a brief review of a supersymmetric matrix model discussed in ref.~\cite{Kuroki:2009yg}. 
The action is 
\be
S = N \tr \left[\frac12 B^2 +iB(\phi^2-\mu^2) +\bar\psi (\phi\psi+\psi\phi)\right],  
\label{S}
\ee
where $B$ and $\phi$ are $N\times N$ hermitian matrices, and $\psi$ and $\bar\psi$ are $N\times N$ 
Grassmann-odd matrices. $S$ is invariant under supersymmetry transformations generated by $Q$ and $\bar{Q}$: 
\be
Q\phi =\psi, \quad Q\psi=0, \quad Q\bar{\psi} =-iB, \quad QB=0, 
\label{QSUSY}
\ee
and 
\be
\bar{Q} \phi = -\bar{\psi}, \quad \bar{Q}\bar{\psi} = 0, \quad 
\bar{Q} \psi = -iB, \quad \bar{Q} B = 0,  
\label{QbarSUSY}
\ee
from which they are nilpotent: $Q^2=\bar{Q}^2=0$. 
The partition function is defined by 
\be
Z  \equiv    \left(-1\right)^{N^2}
\int d^{N^2}B \,d^{N^2}\phi \,d^{N^2}\psi \,d^{N^2}\bar{\psi}\, e^{-S},  
\label{Z}
\ee
where we fix the normalization of the measure as 
\be
\int d^{N^2}\phi \, e^{-N\tr \,(\frac12 \phi^2)} = \int d^{N^2}B \, e^{-N\tr \,(\frac12 B^2)} = 1, \qquad 
(-1)^{N^2} \int \left(d^{N^2}\psi \,d^{N^2}\bar{\psi}\right)\, e^{-N\tr \,(\bar{\psi}\psi)}=1. 
\label{normalization}
\ee
We express the expectation value of a single trace operator $\frac{1}{N}\tr\,\cO$ 
in the $\frac{1}{N}$-expansion as 
\be
\vev{\frac{1}{N}\tr\,\cO} = \vev{\frac{1}{N}\tr\,\cO}_0 + \frac{1}{N^2}\,\vev{\frac{1}{N}\tr\,\cO}_1 + 
\frac{1}{N^4}\,\vev{\frac{1}{N}\tr\,\cO}_2 +\cdots. 
\label{genus_exp}
\ee 
$\vev{\cdot}_h$ is an $N$-independent quantity representing an amplitude over random surfaces 
with handle $h$. 
Similarly, the expansion of the connected correlator of $K$ single-trace operators is 
\be
\vev{\prod_{k=1}^K\frac{1}{N}\tr\,\cO_k}_C = \sum_{h=0}^\infty \frac{1}{N^{2K+2h-2}}\vev{\prod_{k=1}^K\frac{1}{N}\tr\,\cO_k}_{C, h} 
\label{genus_exp_2}
\ee
with $\vev{\cdot}_{C, h}$ $N$-independent. 
 
After integrating out $B$, we find that the scalar potential of $\phi$ is in shape of a double-well: 
$\frac12 (\phi^2-\mu^2)^2$.     
As shown in ref.~\cite{Kuroki:2009yg}, there are infinitely many large-$N$ saddle points 
preserving supersymmetries for $\mu^2>2$. 
The eigenvalue distribution $\rho(x)=\vev{\frac{1}{N}\tr\,\delta(x-\phi)}_0$ has the support  
$\Omega=[-b,-a]\cup [a,b]$ with 
\be
a=\sqrt{-2+\mu^2}, \quad b=\sqrt{2+\mu^2},  
\label{a_b}
\ee
and its explicit form is 
\be
\rho(x) = \left\{\begin{array}{cl} \frac{\nu_+}{\pi}\,x\,\sqrt{(x^2-a^2)(b^2-x^2)} & \qquad (a<x<b) \\
                                   \frac{\nu_-}{\pi}\,|x|\,\sqrt{(x^2-a^2)(b^2-x^2)} & \qquad (-b<x<-a).
                 \end{array}\right. 
\label{rho}                 
\ee 
The filling fraction $\nu_+$ ($\nu_-$) expresses the amount of the eigenvalues 
around the right (left) minimum of the double-well, 
satisfying $\nu_\pm\geq 0$, $\nu_++\nu_-=1$. 
The cases of $(\nu_+, \nu_-) = (1,0)$ and $(0,1)$ reduce the support of $\rho(x)$ to a single interval of 
$[a, b]$ and $[-b,-a]$, respectively.   
The large-$N$ planar free energy $F_0$, which is the leading term of  
\be
F\equiv -\frac{1}{N^2}\ln Z = \sum_{h=0}^\infty \frac{1}{N^{2h}} \,F_h, 
\label{F}
\ee
calculated from (\ref{rho}) vanishes, and all of $\vev{\frac{1}{N}\tr\,B^n}_0$ ($n=1,2,\cdots$) do so.  
It indicates that the solution preserves the supersymmetry. 
Note that the end points of the intervals (\ref{a_b}) do not depend on $\nu_\pm$. 
Because it is not the case of bosonic double-well matrix models~\cite{Cicuta:1986pu,Nishimura:2003gz}, 
it is considered to be a characteristic feature given in supersymmetrizing the double-well matrix model. 
It is easy to see from \eqref{a_b} that the supersymmetric two-cut solution \eqref{rho} ceases to exist 
for $\mu^2<2$. 
In fact, in ref.~\cite{Kuroki:2010au} it is found that when $\mu^2<2$ there is only a vacuum 
without supersymmetry and that we have a third-order phase transition 
between supersymmetric and supersymmetry breaking phase at $\mu^2=2$. 
Hence $\mu^2=2$ is a critical point in our model \eqref{S} and we will elaborate on 
critical behavior of correlation functions in the supersymmetric phase as $\mu^2\rightarrow 2+0$. 

The model are interpreted as the $O(n)$ model on a random lattice with $n=-2$, whose critical behavior is 
described by the $c=-2$ topological gravity~\cite{Kostov:1992pn} 
(the $k=1$ case of $(2,2k-1)$ minimal string theory 
constructed by the one-matrix model~\cite{Brezin:1990rb,Douglas:1989ve,Gross:1989aw}). 
The string susceptibility exponent is $\gamma=-1$. 
The partition function (\ref{Z}) after $B$, $\psi$ and $\bar{\psi}$ integrated out becomes  
a Gaussian one-matrix model by the Nicolai mapping $H=\phi^2-\mu^2$, 
where the integration is over region where $H+\mu^2$ is a {\em positive definite} hermitian matrix, 
not over all the hermitian matrices. 
Ref.~\cite{Gaiotto:2004nz} discusses that since the difference of the integration region has only effects 
which are nonperturbative in $1/N$, the model can be regarded as the standard Gaussian matrix model 
integrated over all the hermitian matrices at each order of genus expansion. 

The Nicolai mapping changes the operators $\frac{1}{N}\tr\,\phi^{2n}$ ($n=1,2,\cdots$) to regular operators 
$\frac{1}{N}\tr\,(H+\mu^2)^{n}$ with respect to $H$. Hence the behavior of their correlators is expected to be 
described by the Gaussian one-matrix model (the $c=-2$ topological gravity) at least perturbatively in $1/N$. 
Actually, we see that $\vev{\frac{1}{N}\tr\,\phi^{2n}}_0$ is a regular 
function of $\mu^2$ in section~\ref{sec:1pt_phi}, 
and that the planar expectation value of the resolvent $\vev{\frac{1}{N}\tr\,\frac{1}{z-\phi^2}}_0$ obeys 
the semi-circle law from (\ref{R2_0}) in appendix~\ref{app:SD}. 
However, the operators $\frac{1}{N}\tr\,\phi^{2n+1}$ ($n=0,1,2,\cdots$) are mapped to 
$\pm\frac{1}{N}\tr\,(H+\mu^2)^{n+1/2}$ that are singular at $H=-\mu^2$. 
They are not observables in the $c=-2$ topological 
gravity, while they are natural observables as well as $\frac{1}{N}\tr\,\phi^{2n}$ 
in the original setting (\ref{Z}). 
In the following sections, we will see that correlation functions among operators~\footnote{
Note that $\tr\,\psi^{2n}=\tr\,\bar{\psi}^{2n}=0$ ($n=1,2,\cdots$) from the cyclicity of the trace and  
the Grassmann nature of $\psi$ and $\bar{\psi}$.}    
\be
\frac{1}{N}\tr\,\phi^{2n+1}, \qquad \frac{1}{N}\tr\,\psi^{2n+1}, \qquad \frac{1}{N}\tr\,\bar{\psi}^{2n+1}
 \qquad (n=0,1,2,\cdots) 
\label{phi_psi_odd} 
\ee
exhibit logarithmic singular behavior of powers of  $\ln(\mu^2-2)$ at the planar topology.   
Note that it is not straightforward to obtain correlation functions of such operators from 
Schwinger-Dyson equations as seen in appendix~\ref{app:SD}.

\section{One-point functions} 
\label{sec:1pt_phi}
\setcounter{equation}{0}
The expectation value of $\frac{1}{N}\tr\,\phi^n$ ($n=1,2,\cdots$) in the large-$N$ limit is expressed 
in term of the hypergeometric function as  
\bea
\vev{\frac{1}{N}\,\tr\,\phi^n}_0 & = & \int_\Omega dx\, x^n \rho(x) 
= (\nu_+\pm \nu_-)\frac{1}{\pi}\int^b_a dx\,x^{n+1}\sqrt{(x^2-a^2)(b^2-x^2)} \nn \\
 & = & (\nu_+\pm \nu_-)\,b^n\,F\left(-\frac{n}{2},\frac32,3;\frac{4}{b^2}\right),   
\label{disk_phi} 
\eea
where the plus (minus) branch of $(\nu_+\pm \nu_-)$ is taken for $n$ even (odd). 
Note that when $n$ is even, the hypergeometric function 
$F\left(-\frac{n}{2},\frac32,3;\frac{4}{b^2}\right)$ is reduced to 
a polynomial of $\mu^2$ giving nonsingular behavior with respect to $\mu^2$:  
\be
\vev{\frac{1}{N}\tr\,\phi^2}_0 = \mu^2, \qquad
\vev{\frac{1}{N}\tr\,\phi^4}_0 = 1+\mu^4, \qquad 
\cdots. 
\label{disk_phi_even}
\ee 

On the other hand, when $n$ is odd, it exhibits logarithmic singular behavior for 
$\omega\equiv \frac{a^2}{4}=\frac{\mu^2-2}{4}\sim +0$: 
\bea
\vev{\frac{1}{N}\tr\,\phi}_0 & = & 
(\nu_+-\nu_-)\left[\frac{64}{15\pi}+\frac{16}{3\pi}\omega +\frac{2}{\pi}\omega^2\ln \omega
+\cO(\omega^2)\right], \nn\\
\vev{\frac{1}{N}\tr\,\phi^3}_0 & = & 
(\nu_+-\nu_-)\left[\frac{1024}{105\pi}+\frac{128}{5\pi}\omega 
+ \frac{16}{\pi} \omega^2+\frac{4}{\pi}\omega^3\ln \omega 
+\cO(\omega^3)\right], \nn\\
\vev{\frac{1}{N}\tr\,\phi^5}_0 & = & (\nu_+-\nu_-)\left[\frac{8192}{315\pi}+\frac{2048}{21\pi}\omega 
+ \frac{128}{\pi} \omega^2+\frac{160}{3\pi}\omega^3+\frac{10}{\pi}\omega^4\ln \omega 
+\cO(\omega^4)\right], \nn\\
 & \cdots  & .  
\label{disk_phi_several} 
\eea  
In general, for $k=0,1,2,\cdots$, 
\be
\left.\vev{\frac{1}{N}\tr\,\phi^{2k+1}}_0 \right|_{\rm sing.} 
= (\nu_+-\nu_-)\left[\frac{2^{k+2}}{\pi}\frac{(2k+1)!!}{(k+2)!}\,
\omega^{k+2} \ln \omega +(\mbox{less singular})\right], 
\label{disk_phi_general}
\ee
where $|_{\rm sing.}$ stands for removing entire functions (polynomials) of $\omega$. 

The singular term of $\ln \omega$ becomes dominant after taking $\mu^2$-derivatives ($k+2$) times, 
namely
\bea
\frac{\partial^{k+2}}{\partial (\mu^2)^{k+2}}\vev{\frac{1}{N}\tr\,\phi^{2k+1}}_0 & = & 
\vev{\frac{1}{N}\tr\,\phi^{2k+1}\,\left(\frac{1}{N}\,\tr\, iB\right)^{k+2}}_{C,0} \nn \\
 & = &  
(\nu_+-\nu_-)\left[\frac{(2k+1)!!}{2^{k+2}\,\pi}\,\ln \omega + \cO(\omega^0, \,\omega\ln\omega)\right]. 
\eea

\section{Two-point functions of bosons} 
\label{sec:2pt_phi}
\setcounter{equation}{0}
In computing higher-point correlators $\vev{\prod_{i=1}^K\frac{1}{N}\tr\,\phi^{n_i}}_{C,0}$ 
at the vacuum with general filling fractions $(\nu_+, \nu_-)$, it is useful 
to reduce them to those at the vacuum with $(\nu_+, \nu_-)=(1,0)$.  
It is discussed in appendix~\ref{app:ff}, where 
\be
\vev{\prod_{i=1}^K\frac{1}{N}\tr\,\phi^{n_i}}^{(\nu_+,\nu_-)}_{C,0}= (\nu_+-\nu_-)^\sharp 
\vev{\prod_{i=1}^K\frac{1}{N}\tr\,\phi^{n_i}}^{(1,0)}_{C,0}
\label{nu_pm_to_10}
\ee
is shown up to $K=3$. Here, the superscripts $(\nu_+,\nu_-)$ and $(1,0)$ are put to clarify the filling 
fractions of the vacua at which    
the expectation values are evaluated, and $\sharp$ counts 
the number of odd integers in $\{n_1, \cdots, n_K\}$. 
 
\subsection{Eigenvalue distribution with source}
In order to obtain higher-point correlators of $\frac{1}{N}\tr\,\phi^p$ ($p=1,2,\cdots$), 
we introduce 
source terms $\sum_{p=1}^\infty j_p \tr \,\phi^p$ to the partition function:  
\be
Z_{j_k} = \int d^{N^2}\phi\, e^{-N\tr\left[\frac12(\phi^2-\mu^2)^2 -\sum_{p=1}^\infty j_p\phi^p\right]}\,
\det\left(\phi\otimes\id + \id \otimes\phi\right). 
\label{Zj}
\ee
In the large-$N$ limit, the eigenvalue distribution $\rho_j(x)$ satisfies the saddle point equation  
\be
\int dy \,\rho_j(y)\, \left({\rm P}\frac{1}{x-y} + {\rm P}\frac{1}{x+y}\right) 
= x^3-\mu^2 x -\sum_{p=1}^\infty \frac{pj_p}{2}\,x^{p-1}. 
\label{SPE0_j}
\ee
Let us consider the case of the filling fractions $(1,0)$~\footnote{
Curiously, the saddle point equation (\ref{SPE0_j}) is sometimes 
not consistent for $j_p\neq 0$ with $p$ odd, 
when the support of $\rho_j(x)$ consists of two intervals 
$\Omega_j=[-b_j',-a_j']\cup [a_j,b_j]$ ($0<a_j<b_j$, $0<a_j'<b_j'$). 
If there is a point $x\in \Omega_j$ such that also $-x\in\Omega_j$, 
eq.~\eqref{SPE0_j} for $x$ and that for $-x$ contradict each other, 
because the l.h.s. of (\ref{SPE0_j}) 
is odd for $x\to -x$, while the r.h.s. is not due to the source $j_p$ ($p$: odd). 
The l.h.s. tells that the eigenvalue at the position $x$ feels repulsive forces from the other eigenvalues 
as well as from the mirror images of all the eigenvalues. 
Since it becomes an odd function of $x$, 
total force is not balanced unless the matrix model potential is even. 
Thus, 
the deformation by the source $j_p$ ($p$: odd) makes the general two-cut solution (\ref{rho}) unstable, 
and would drastically change the support of the eigenvalue distribution. 
\\
To avoid this difficulty, we first solve (\ref{SPE0_j}) for the filling fractions $(1,0)$, 
and then obtain the amplitudes for general $(\nu_+,\nu_-)$ by using (\ref{nu_pm_to_10}).   
}   
with the support of $\rho_j(x)$ $[a_j,b_j]$ ($0<a_j<b_j$). 
We change variables as 
\be
x^2=A+B\xi, \qquad y^2=A+B\eta  
\qquad \mbox{with} \qquad 
A\equiv \frac{a_j^2+b_j^2}{2}, \qquad B\equiv \frac{b_j^2-a_j^2}{2},
\ee
and put 
$\tilde{\rho}(\eta)\equiv \frac{B}{2y}\rho_j(y)$,  
to simplify (\ref{SPE0_j}) as 
\be
\frac{1}{B}\int^1_{-1}d\eta\,\tilde{\rho}(\eta) \,{\rm P}\frac{1}{\xi-\eta} 
= \frac12(A-\mu^2+B\xi)-\sum_{p=1}^\infty\frac{pj_p}{4}\,(A+B\xi)^{\frac{p}{2}-1}
\label{SPE2_j}
\ee
for $\xi\in[-1,1]$, where $\tilde{\rho}$ is normalized by  
$\int ^1_{-1}d\eta \tilde{\rho}(\eta)=1$. 

We act $\int^1_{-1}d\xi\,\sqrt{1-\xi^2}\,{\rm P}\frac{1}{\zeta-\xi}$ to both sides of (\ref{SPE2_j}), 
and apply the formula (\ref{app:formula}) in appendix~\ref{app:formulas}. Then we obtain
\bea
\tilde{\rho}(\zeta) & = & \frac{1}{2\pi}\frac{1}{\sqrt{1-\zeta^2}}\left[2-(A-\mu^2)B\zeta 
-B^2\left(\zeta^2-\frac12\right) \right. \nn \\
& & \hspace{24mm} \left. +\sum_{p=1}^\infty pj_p\,\frac{B}{2\pi}\int^1_{-1}d\xi\, \sqrt{1-\xi^2}\,{\rm P}\frac{1}{\zeta-\xi}
\,(A+B\xi)^{\frac{p}{2}-1}\right]
\label{rho_tilde}
\eea 
after using (\ref{app:GMM0}) and (\ref{app:GMM1}). 
The condition $\tilde{\rho}(\zeta=\pm 1)=0$ determines $A$ and $B$ as 
\bea
A & = & \mu^2 +\sum_{p=1}^\infty j_p\,\frac{p}{2} (A+B)^{\frac{p}{2}-1} F\left(-\frac{p}{2}+1, \frac12, 1; \frac{2B}{A+B}\right), 
\label{A_eq} \\
B & = & 2\left[1+\sum_{p=1}^\infty\frac{j_p}{4}\,\frac{p}{2}\left(\frac{p}{2}-1\right)B^2(A+B)^{\frac{p}{2}-2} 
F\left(-\frac{p}{2}+2, \frac32, 3;\frac{2B}{A+B}\right)\right]^{1/2},  
\label{B_eq}
\eea
from which $A$ and $B$ are obtained iteratively with respect to $\{j_p\}$. 
Up to the first order of $\{j_p\}$, 
\bea
A& = & \mu^2+\sum_{p=1}^\infty j_p\,\frac{p}{2}(2+\mu^2)^{\frac{p}{2}-1}
F\left(-\frac{p}{2}+1,\frac12,1;\frac{4}{2+\mu^2}\right)  + \cO(j^2),  
\label{A_sol}\\
B & = & 2+\sum_{p=1}^\infty j_p\,\frac{p}{2}\left(\frac{p}{2}-1\right)(2+\mu^2)^{\frac{p}{2}-2} 
F\left(-\frac{p}{2}+2,\frac32,3;\frac{4}{2+\mu^2}\right) + \cO(j^2),  
\label{B_sol}
\eea
where $\cO(j^2)$ means a quantity of the quadratic order of $\{j_p\}$.  

\subsection{Two-point amplitudes $\vev{\frac{1}{N}\tr\,\phi^p\,\frac{1}{N}\tr\,\phi^q}_{C,0}$}
\label{sec:cyl_p_q}
Let us express the planar expectation value of $\cO$ 
under the partition function with the source terms (\ref{Zj}) as 
$\vev{\cO}_0^{(j)}$. The cylinder amplitude at the vacuum with the filling fractions $(1,0)$ is given as 
\bea
\vev{\frac{1}{N}\tr\,\phi^p\,\frac{1}{N}\tr\,\phi^q}^{(1,0)}_{C,0} & = & \left.
\frac{\partial}{\partial j_p}\vev{\frac{1}{N}\tr\,\phi^q}_0^{(j)}\right|_{\{j_p\}=0} \nn \\
& = & \left.
\frac{\partial}{\partial j_p}\int^1_{-1}d\zeta\,(A+B\zeta)^{\frac{q}{2}}\tilde{\rho}(\zeta)\right|_{\{j_p\}=0}. 
\label{cyl0}
\eea
Plugging (\ref{rho_tilde}), (\ref{A_sol}) and (\ref{B_sol}) into (\ref{cyl0}) leads to 
\bea
\vev{\frac{1}{N}\tr\,\phi^p\,\frac{1}{N}\tr\,\phi^q}^{(1,0)}_{C,0}
& = & \left.\frac{1}{\pi}\int^1_{-1}d\zeta\,\partial_{\zeta}\left[(\mu^2+2\zeta)^{\frac{q}{2}}\sqrt{1-\zeta^2} 
\left(\partial_{j_p}A+\partial_{j_p}B\,\zeta\right)\right]\right|_{\{ j_p\}=0} \nn \\
& & + \frac{p}{2\pi^2}\int^1_{-1}d\zeta\,\frac{(\mu^2+2\zeta)^{\frac{q}{2}}}{\sqrt{1-\zeta^2}}
\int^1_{-1}d\xi\,\sqrt{1-\xi^2}\,(\mu^2+2\xi)^{\frac{p}{2}-1}\nn \\
& & \hspace{55mm} \times{\rm P}\frac{1}{\zeta-\xi}. 
\label{cyl1_10}
\eea
Note that the linear-order contributions to $A$ and $B$ with respect to $\{ j_p\}$ appear in 
the first term, but it is a total derivative term and vanishes. 
Thus, from (\ref{nu_pm_to_10}) the amplitude at general filling fractions is 
\bea
& & \vev{\frac{1}{N}\tr\,\phi^p\,\frac{1}{N}\tr\,\phi^q}^{(\nu_+,\nu_-)}_{C,0} \nn \\
& & = (\nu_+-\nu_-)^\sharp \frac{p}{2\pi^2}\int^1_{-1}d\zeta\,
\frac{(\mu^2+2\zeta)^{\frac{q}{2}}}{\sqrt{1-\zeta^2}}
\int^1_{-1}d\xi\,\sqrt{1-\xi^2}\,(\mu^2+2\xi)^{\frac{p}{2}-1}\,{\rm P}\frac{1}{\zeta-\xi},  
\label{cyl1_nu_pm}
\eea
where $\sharp=2$ for $p$ and $q$ odd, $\sharp=0$ for $p$ and $q$ even, and $\sharp=1$ for one of 
$p$ and $q$ odd.  
By integrating by parts (\ref{cyl1_nu_pm}) with respect to $\xi$, 
we obtain a symmetric expression for $p$ and $q$: 
\bea
 & & \vev{\frac{1}{N}\tr\,\phi^p\,\frac{1}{N}\tr\,\phi^q}^{(\nu_+,\nu_-)}_{C,0} \nn \\
& & = (\nu_+-\nu_-)^\sharp 
\frac{1}{2\pi^2}\int^1_{-1}d\zeta\,\frac{(\mu^2+2\zeta)^{\frac{q}{2}}}{\sqrt{1-\zeta^2}}
\int^1_{-1}d\xi\,\frac{(\mu^2+2\xi)^{\frac{p}{2}}}{\sqrt{1-\xi^2}}
\,{\rm P}\frac{1}{(\zeta-\xi)^2}\,(\zeta\xi-1).  
\label{cyl2_nu_pm}
\eea

\subsubsection{Case of $p$ and $q$ odd}

Let us compute further (\ref{cyl1_nu_pm}) for case of $p$ and $q$ odd.  
We can put $p=2k+1$, $q=2\ell+1$ with $k\leq \ell$, i.e. $\ell=k+m$ ($m=0,1,2,\cdots$),  
without loss of generality. 
Symmetrizing (\ref{cyl1_nu_pm}) with respect to $\zeta$ and $\xi$ leads to 
\bea
 & & \vev{\frac{1}{N}\tr\,\phi^{2k+1}\,\frac{1}{N}\tr\,\phi^{2\ell+1}}_{C,0} \nn \\
& & = (\nu_+-\nu_-)^2\frac{2k+1}{4\pi^2}\int^1_{-1}d\zeta\,\frac{(\mu^2+2\zeta)^{k-1/2}}{\sqrt{1-\zeta^2}}
\int^1_{-1}d\xi\,\frac{(\mu^2+2\xi)^{k-1/2}}{\sqrt{1-\xi^2}} \,{\rm P}\frac{1}{\zeta-\xi} \nn \\
& & \hspace{17mm}\times \left[(\mu^2+2\zeta)^{m+1}(1-\xi^2)-(\mu^2+2\xi)^{m+1}(1-\zeta^2)\right]. 
\eea
Here and in what follows, we omit the superscript $(\nu_+,\nu_-)$ when it does not cause any confusion. 
We express the amplitude in terms of 
\be
I_{p-\frac12}\equiv \int^1_{-1}dx\,\frac{(\mu^2+2x)^{p-\frac12}}{\sqrt{1-x^2}}
= \pi(4(1+\omega))^{p-\frac12}F\left(\frac12-p,\frac12,1;\frac{1}{1+\omega}\right)  \qquad (p=0,1,2,\cdots)
\label{I_p}
\ee
as 
\bea
 & & \vev{\frac{1}{N}\tr\,\phi^{2k+1}\,\frac{1}{N}\tr\,\phi^{2\ell+1}}_{C,0} \nn \\
& & = (\nu_+-\nu_-)^2\frac{2k+1}{8\pi^2}\left[\sum_{p=1}^mI_{p+k-\frac12}\left\{-I_{m-p+k+\frac32}
+2\mu^2I_{m-p+k+\frac12}+(4-\mu^4)I_{m-p+k-\frac12}\right\} \right. \nn \\
& & \hspace{39mm} \left.+I_{k+\frac12}I_{m+k+\frac12}+(4-\mu^4)I_{k-\frac12}I_{m+k-\frac12}\frac{}{}\right].
\label{phi_odd_cyl}
\eea

Note that 
$
I_{p-\frac12} 
$ 
yields logarithmic singular behavior ($\sim \omega^p\ln \omega$) for $\omega\sim +0$. 
For $p\neq 0$, however $I_{p-\frac12}$ has a finite limit as $\omega\to +0$. In fact we find 
\bea
I_{p-\frac12} & = & -2^{p-1}\,\frac{(2p-1)!!}{p!}\,\omega^p\ln\omega +\cO(\omega^{p+1}\ln \omega, \omega^p) 
\nn \\
& & +\left(2^{3p-1}\,\frac{(p-1)!}{(2p-1)!!}+\cO(\omega)\right)(1-\delta_{p,0})
\label{Iexpansion}
\eea
with $(-1)!!=1$. 

The explicit form of the first several amplitudes are 
\bea
\left.\vev{\frac{1}{N}\tr\,\phi\,\frac{1}{N}\tr\,\phi}_{C,0}\right|_{\rm sing.} 
& = & (\nu_+-\nu_-)^2\left[-\frac{1}{2\pi^2}\,\omega(\ln\omega)^2 +\cO(\omega\ln\omega)\right], \nn \\
\left.\vev{\frac{1}{N}\tr\,\phi\,\frac{1}{N}\tr\,\phi^3}_{C,0}\right|_{\rm sing.} 
 & = &  (\nu_+-\nu_-)^2\left[\frac{4}{\pi^2}\,\omega\ln\omega 
-\frac{3}{2\pi^2}\,\omega^2(\ln\omega)^2 +\cO(\omega^2\ln\omega)\right], \nn \\
\left.\vev{\frac{1}{N}\tr\,\phi^3\,\frac{1}{N}\tr\,\phi^3}_{C,0}\right|_{\rm sing.} 
 & = & (\nu_+-\nu_-)^2\left[\frac{24}{\pi^2}\,\omega^2\ln\omega 
-\frac{6}{\pi^2}\omega^3(\ln\omega)^2 +\cO(\omega^3\ln\omega)\right], \nn \\
\left.\vev{\frac{1}{N}\tr\,\phi\,\frac{1}{N}\tr\,\phi^5}_{C,0}\right|_{\rm sing.} 
 & = & (\nu_+-\nu_-)^2\left[\frac{32}{3\pi^2}\,\omega\ln\omega 
+\frac{32}{\pi^2}\,\omega^2\ln\omega 
-\frac{5}{\pi^2}\,\omega^3(\ln\omega)^2 \right. \nn \\
 & & \hspace{21mm}\left. \frac{}{}+ \cO(\omega^3\ln\omega)\right], \nn 
\\
%
\left.\vev{\frac{1}{N}\tr\,\phi^3\,\frac{1}{N}\tr\,\phi^5}_{C,0}\right|_{\rm sing.} 
 & = & (\nu_+-\nu_-)^2\left[\frac{32}{\pi^2}\,\omega^2\ln\omega 
+\frac{144}{\pi^2}\,\omega^3\ln\omega 
 -\frac{45}{2\pi^2}\,\omega^4(\ln\omega)^2 \right.\nn \\
& & \hspace{21mm}\left.\frac{}{}+\cO(\omega^4\ln\omega)\right], \nn 
\\
%
\left.\vev{\frac{1}{N}\tr\,\phi^5\,\frac{1}{N}\tr\,\phi^5}_{C,0}\right|_{\rm sing.} 
 & = & (\nu_+-\nu_-)^2\left[\frac{640}{3\pi^2}\,\omega^3\ln\omega 
+\frac{720}{\pi^2}\,\omega^4\ln\omega -\frac{90}{\pi^2}\,\omega^5(\ln\omega)^2 \right.\nn \\
& & \hspace{21mm}\left.\frac{}{}+\cO(\omega^5\ln\omega)\right]. 
%
\label{cyl_odd_several}
\eea

A characteristic feature of the amplitudes is singular behavior of $(\ln \omega)^2$ multiplied by 
powers of $\omega$. 
The leading singular term of $(\ln\omega)^2$ behavior 
in $\vev{\frac{1}{N}\tr\,\phi^{2k+1}\,\frac{1}{N}\tr\,\phi^{2\ell+1}}_{C,0}$ with $\ell=k+m$ ($m=0,1,2,\cdots$) takes the form:  
\bea
& & \left.\vev{\frac{1}{N}\tr\,\phi^{2k+1}\,\frac{1}{N}\tr\,\phi^{2\ell+1}}_{C,0}\right|_{\mbox{$(\ln\omega)^2$-leading}} \nn \\
& & =-(\nu_+-\nu_-)^2\frac{2k+1}{4\pi^2}\,2^{2k+m}\left[\sum_{p=1}^m\frac{(2p+2k-1)!!}{(p+k)!} 
\frac{(2m-2p+2k-1)!!}{(m-p+k+1)!} \right.\nn \\
& & \hspace{52mm} \left. +2\,\frac{(2k-1)!!}{k!}\frac{(2m+2k-1)!!}{(m+k)!}\right] 
\omega^{2k+m+1}(\ln\omega)^2. \nn \\
& & 
\label{odd_cyl_ln2}
\eea

\subsubsection{Case of $p$ odd and $q$ even}
\label{sec:cyl_odd_even}
Let us compute the amplitude for case of $p$ odd ($p=2k+1$) and $q$ even ($q=2\ell$) 
in \eqref{cyl1_nu_pm}: 
\bea 
 & & \vev{\frac{1}{N}\tr\,\phi^{2k+1}\,\frac{1}{N}\tr\,\phi^{2\ell}}^{(\nu_+,\nu_-)}_{C,0} \nn \\
& & = (\nu_+-\nu_-) \,\frac{2k+1}{2\pi^2}\int^1_{-1}d\zeta\,
\frac{(\mu^2+2\zeta)^{\ell}}{\sqrt{1-\zeta^2}}
\int^1_{-1}d\xi\,\sqrt{1-\xi^2}\,(\mu^2+2\xi)^{k-\frac12}\,{\rm P}\frac{1}{\zeta-\xi}. \nn \\ 
\label{cyl_even3}
\eea   
Use of 
\be
(\mu^2+2\zeta)^{\ell}=\sum_{r=0}^\ell \begin{pmatrix} \ell \\ r\end{pmatrix} (\mu^2+2\xi)^{\ell-r}\,
2^r(\zeta-\xi)^r
\ee
and (\ref{app:formula3}) gives 
\bea
 & & \int^1_{-1}d\xi\,\frac{(\mu^2+2\zeta)^{\ell}}{\sqrt{1-\zeta^2}}\,{\rm P}\frac{1}{\zeta-\xi} \nn \\
 & & = -\pi\sum_{r=1}^\ell\sum_{m=0}^{\left[\frac{r-1}{2}\right]}\frac{1}{r}\frac{\ell!}{(\ell-r)!(r-1-2m)!} 
\frac{(-2)^{r-2m}}{(m!)^2}\,(\mu^2+2\xi)^{\ell-r}\xi^{r-1-2m} 
\eea 
with $[x]$ the greatest integer not exceeding $x$. 
Plugging this into (\ref{cyl_even3}), we have
\bea
 & & \vev{\frac{1}{N}\tr\,\phi^{2k+1}\,\frac{1}{N}\tr\,\phi^{2\ell}}_{C,0} \nn \\
 & & = (\nu_+-\nu_-) 
\frac{1}{\pi}(2k+1)\sum_{r=1}^\ell\sum_{m=0}^{\left[\frac{r-1}{2}\right]}\sum_{s=0}^{r-1-2m} 
 \frac{1}{r}\frac{\ell!}{(\ell-r)!s!(r-1-2m-s)!} 
 \frac{(-2)^{r-2m+s+1}}{(m!)^2}\nn \\
& & \hspace{17mm}\times (4(1+\omega))^{k+\ell-r-\frac12}
 B\left(\frac32, s+\frac32\right) F\left(-k-\ell+r+\frac12, s+\frac32, s+3; \frac{1}{1+\omega}\right), \nn \\
 & & 
 \label{cyl_even4}
\eea 
from which 
logarithmic singular behavior of the amplitude is seen as 
\bea
\left. \vev{\frac{1}{N}\tr\,\phi^{2k+1}\,\frac{1}{N}\tr\,\phi^{2\ell}}_{C,0} \right|_{\rm sing.}
 & = & (\nu_+-\nu_-)\left[
\frac{2^{k+\ell}}{2\pi}\frac{(2k+1)!!}{(k+1)!}\frac{(2\ell-3)!!}{(\ell-1)!} \,\omega^{k+1}\ln\omega \right.
\nn \\
& & \hspace{19mm} \left. \frac{}{} +(\mbox{less singular})\right]. 
 \label{cyl_even_sing}
\eea

The first several amplitudes are explicitly given as   
\bea
\left.\vev{\frac{1}{N}\tr\,\phi\,\frac{1}{N}\tr\,\phi^2}_{C,0}\right|_{\rm sing.}
& = &  (\nu_+-\nu_-)\left[\frac{1}{\pi}\omega\ln\omega 
+ \frac{3}{8\pi}\,\omega^2\ln\omega + \cO(\omega^3\ln \omega)\right], \nn \\
\left.\vev{\frac{1}{N}\tr\,\phi\,\frac{1}{N}\tr\,\phi^4}_{C,0}\right|_{\rm sing.}
& = & (\nu_+-\nu_-)\left[ \frac{2}{\pi}\,\omega\ln\omega 
+ \frac{23}{4\pi}\,\omega^2\ln\omega +\cO(\omega^3\ln\omega)\right], \nn \\
\left.\vev{\frac{1}{N}\tr\,\phi^3\,\frac{1}{N}\tr\,\phi^2}_{C,0}\right|_{\rm sing.}
& = & (\nu_+-\nu_-)\left[\frac{3}{\pi}\,\omega^2\ln\omega 
+ \frac{3}{4\pi}\,\omega^3\ln\omega 
+ \cO(\omega^4\ln\omega)\right], \nn \\
\left.\vev{\frac{1}{N}\tr\,\phi^3\,\frac{1}{N}\tr\,\phi^4}_{C,0}\right|_{\rm sing.}
 & = & (\nu_+-\nu_-)\left[ \frac{6}{\pi}\,\omega^2\ln\omega 
+ \frac{39}{2\pi}\,\omega^3\ln\omega + \cO(\omega^4\ln\omega) \right], \nn 
\\
%
\left.\vev{\frac{1}{N}\tr\,\phi^5\,\frac{1}{N}\tr\,\phi^2}_{C,0}\right|_{\rm sing.} 
& = & (\nu_+-\nu_-)\left[\frac{10}{\pi}\,\omega^3\ln\omega 
+ \frac{15}{8\pi}\,\omega^4\ln \omega  +\cO(\omega^5\ln\omega)\right], \nn 
\\
%
\left.\vev{\frac{1}{N}\tr\,\phi^5\,\frac{1}{N}\tr\,\phi^4}_{C,0}\right|_{\rm sing.}  
 & = & (\nu_+-\nu_-)\left[\frac{20}{\pi}\,\omega^3\ln\omega
+ \frac{275}{4\pi}\,\omega^4\ln\omega +\cO(\omega^5\ln\omega)\right]. \nn \\
& & 
\label{cyl_even_several}
\eea

\subsubsection{Case of $p$ and $q$ even}
\label{sec:even_even}
The amplitude (\ref{cyl1_nu_pm}) becomes a polynomial of $\omega$ for case of $p$ and $q$ even. 
In this case \eqref{cyl1_nu_pm} can be derived independently by (\ref{evencyl}) 
from which we read off 
\be
\vev{\frac{1}{N}\tr\,\phi^2\, \frac{1}{N}\tr\,\phi^2}_{C,0} =  1, \qquad 
\vev{\frac{1}{N}\tr\,\phi^2\, \frac{1}{N}\tr\,\phi^4}_{C,0} =  4(1+2\,\omega), 
\cdots. 
\ee

\subsection{Operator mixing}
\label{sec:mixing_phi}
In this subsection, we discuss operator mixing which makes the logarithmic singular behavior 
of the cylinder amplitudes more transparent. 
Eq.~\eqref{cyl_even_sing} indicates that the leading logarithmic behavior $\omega^{k+1}\ln\omega$ 
is brought by the operator of an odd power $\frac{1}{N}\tr\,\phi^{2k+1}$ 
since it does not have $\ell$-dependence. 
(The operator of an even power $\frac{1}{N}\tr\,\phi^{2\ell}$ 
just affects the coefficient of the $\omega^{k+1}\ln\omega$ term.) 
On the other hand, from \eqref{cyl_odd_several} 
we find that in the cylinder amplitude of operators of odd powers of $\phi$, 
the $(\ln\omega)^2$ terms of our interest are in general less singular than some $\ln\omega$ ones.  
These observation suggests that 
we can construct a set of new operators by mixing each operator of an odd power 
and operators of lower even powers in such a way that the leading singular behavior 
in the cylinder amplitudes of the new operators will be given by $(\ln\omega)^2$ terms. 
Note that operators of even powers we add do not affect $(\ln\omega)^2$ terms at all, 
because cylinder amplitudes involving even-power operators do not have a singularity of $(\ln\omega)^2$, 
as seen from \eqref{cyl_even_sing} and 
section~\ref{sec:even_even}.

\subsubsection{Explicit construction}
It is easy to check that this prescription works at least lower orders by explicit construction. 
Since the most singular term in $\vev{\frac{1}{N}\tr\,\phi\,\frac{1}{N}\tr\,\phi}_{C,0}$ 
is given by $\omega (\ln\omega)^2$ as in \eqref{cyl_odd_several}, we do not have to do anything. 
Let us set 
\begin{align}
\Phi_1=\frac1N\tr\,\phi.
\end{align}
Then \eqref{cyl_odd_several} tells us that $\vev{\Phi_1\frac1N\tr\,\phi^3}$ has the $\omega\ln\omega$ term 
which is larger than $\omega^2(\ln\omega)^2$. Thus we combine $\frac1N\tr\,\phi^3$ with 
an operator of the power which is even and smaller than $\phi^3$ as 
\begin{align}
\Phi_3=\frac1N\tr\,\phi^3+(\nu_+-\nu_-)\,c\frac1N\tr\,\phi^2,
\label{Phi3mixing}
\end{align}
where $c$ is a constant fixed by imposing that singular terms 
of $\vev{\Phi_1\Phi_3}$ start from $\omega^2(\ln\omega)^2$. 
{}From \eqref{cyl_odd_several} and \eqref{cyl_even_several}, $c$ 
can be determined as $c=-\frac{4}{\pi}$ so that 
\begin{align}
\left.\vev{\Phi_1\Phi_3}_{C,0}\right|_{\text{sing.}}=
(\nu_+-\nu_-)^2\left[-\frac{3}{2\pi^2}\,\omega^2(\ln\omega)^2+\text{(less singular)}\right].
\label{Phi1Phi3}
\end{align}
It is interesting that once we fix $c=-\frac{4}{\pi}$, 
it also absorbs the unwanted $\omega^2\ln\omega$ term in $\vev{\frac1N\tr\,\phi^3\frac1N\tr\,\phi^3}_{C,0}$ 
in \eqref{cyl_odd_several} as 
\begin{align}
\left.\vev{\Phi_3\Phi_3}_{C,0}\right|_{\text{sing.}}=
(\nu_+-\nu_-)^2\left[-\frac{6}{\pi^2}\,\omega^3(\ln\omega)^2+\text{(less singular)}\right].
\label{Phi3Phi3}
\end{align}

Likewise, we assume 
\begin{align}
\Phi_5=\frac1N\tr\,\phi^5+(\nu_+-\nu_-)\,\alpha\frac1N\tr\,\phi^4+(\nu_+-\nu_-)\,\beta\frac1N\tr\,\phi^2,
\end{align}
and fix $\alpha$ and $\beta$ so that the leading singular term in $\vev{\Phi_k\Phi_5}_{C,0}$ ($k=1,3,5$) 
will contain $(\ln\omega)^2$. 
It turns out to be possible, if we allow $\omega$-dependence 
of $\alpha$ and $\beta$ as polynomials of $\omega$ of lower degree (entire functions of $\omega$) as 
\begin{align}
&\Phi_1=\frac1N\tr\,\phi, \label{Phi1}\\
&\Phi_3=\frac1N\tr\,\phi^3
+(\nu_+-\nu_-)\left(\alpha_{3,2}^{(0)}+\alpha_{3,2}^{(1)}\,\omega+{\cal O}(\omega^2)\right)\frac1N\tr\,\phi^2, 
\label{Phi3}\\
&\Phi_5=\frac1N\tr\,\phi^5+(\nu_+-\nu_-)\left(\alpha_{5,4}^{(0)}+\alpha_{5,4}^{(1)}\,\omega
+{\cal O}(\omega^2)\right)\frac1N\tr\,\phi^4 \nn \\
& \hspace{23mm} 
+(\nu_+-\nu_-)\left(\alpha_{5,2}^{(0)}+\alpha_{5,2}^{(1)}\,\omega+{\cal O}(\omega^2)\right)\frac1N\tr\,\phi^2 
\label{Phi5}
\end{align}
with $\alpha_{3,2}^{(0)}=-\frac{4}{\pi}$. 
Notice that the even-power operators do not affect the singular behavior of the one-point function 
in (\ref{disk_phi_general}): 
\be
\left.\vev{\Phi_{2k+1}}_0\right|_{\rm sing.} = \left.\vev{\frac{1}{N}\tr\,\phi^{2k+1}}_0\right|_{\rm sing.}, 
\ee
and that inclusion of $\alpha_{3,2}^{(1)}\,\omega + \cO(\omega^2)$ in \eqref{Phi3} does not spoil 
\eqref{Phi1Phi3} and 
\eqref{Phi3Phi3} because it only affects higher order, hence less singular terms. 
Then vanishing of $\omega\ln\omega$ term in $\vev{\Phi_1\Phi_5}_{C,0}$, of $\omega^2\ln\omega$ term in 
$\vev{\Phi_3\Phi_5}_{C,0}$, and of $\omega^3\ln\omega$ term in $\vev{\Phi_5\Phi_5}_{C,0}$ 
lead to exactly the same equation:  
\be
0=\frac{32}{3\pi}+2\alpha_{5,4}^{(0)}+\alpha_{5,2}^{(0)}.  
\label{cond15_1st} 
\ee
Furthermore requirement for $\omega^2\ln\omega$ term in $\vev{\Phi_1\Phi_5}_{C,0}$, 
$\omega^3\ln\omega$ term in $\vev{\Phi_3\Phi_5}_{C,0}$ 
and $\omega^4\ln\omega$ term in $\vev{\Phi_5\Phi_5}_{C,0}$ to vanish give 
two independent equations (it turns out that all of the three are not independent): 
\begin{align}
0&=\frac{32}{\pi}+\frac{23}{4}\alpha_{5,4}^{(0)}+\frac{3}{8}\alpha_{5,2}^{(0)}
+2\alpha_{5,4}^{(1)}+\alpha_{5,2}^{(1)}, 
\label{cond15} \\
0&=\frac{104}{\pi}+\frac{39}{2}\alpha_{5,4}^{(0)}+\frac{3}{4}\alpha_{5,2}^{(0)}
+6\alpha_{5,4}^{(1)}+3\alpha_{5,2}^{(1)}.
\label{cond35}
\end{align}
%
%
{}From these equations, we see that the linear combinations 
\begin{align}
&\Phi_1=\frac1N\tr\,\phi, \nn\\
&\Phi_3=\frac1N\tr\,\phi^3
-(\nu_+-\nu_-)\,\frac{4}{\pi}\left(1+\bar{\alpha}_{3,2}^{(1)}\,\omega +{\cal O}(\omega^2)\right)\frac1N\tr\,\phi^2, \nn\\
&\Phi_5=\frac1N\tr\,\phi^5
-(\nu_+-\nu_-)\,\frac{4}{\pi}\left(1+\bar{\alpha}_{5,4}^{(1)}\,\omega +{\cal O}(\omega^2)\right)\frac1N\tr\,\phi^4 \nn \\
& \hspace{23mm} -(\nu_+-\nu_-)\,\frac{8}{3\pi}\left(1+3(1-\bar{\alpha}_{5,4}^{(1)})\,\omega 
+{\cal O}(\omega^2)\right)\frac1N\tr\,\phi^2
\label{Phi}
\end{align}
with $\bar{\alpha}_{3,2}^{(1)}\equiv \alpha_{3,2}^{(1)}/\alpha_{3,2}^{(0)}$, 
$\bar{\alpha}_{5,4}^{(1)}\equiv \alpha_{5,4}^{(1)}/\alpha_{5,4}^{(0)}$ give the behavior 
\be
\left.\vev{\Phi_{2k+1}\,\Phi_{2\ell+1}}_{C, 0}\right|_{\rm sing.} 
= (\nu_+-\nu_-)^2 \left[u_{k,\ell}\,\omega^{k+\ell+1}(\ln\omega)^2
+ (\mbox{less singular})\right]
\label{Phi_odd_cyl}
\ee
for $k, \ell = 0, 1, 2$. $u_{k, \ell}$ is a constant equal to the coefficient of 
the $\omega^{k+\ell+1}(\ln \omega)^2$ term in 
$\vev{\frac{1}{N}\tr\,\phi^{2k+1}\,\frac{1}{N}\tr\,\phi^{2\ell+1}}_{C,0}$. 
$\bar{\alpha}_{3,2}^{(1)}$ and $\bar{\alpha}_{5,4}^{(1)}$ remain undetermined from the requirements 
with respect to $k, \ell=0,1,2$. 
(They could be fixed by considering amplitudes for higher $k$ and $\ell$.) 
%

\subsubsection{General structure of operator mixing}

As we have seen above, the equations for $\alpha_{5,2}^{(0)}, \cdots, \alpha_{5,4}^{(1)}$ 
from the requirement in $\vev{\Phi_1\,\Phi_5}_{C,0}$, $\vev{\Phi_3\,\Phi_5}_{C,0}$ 
and $\vev{\Phi_5\,\Phi_5}_{C,0}$ 
are highly degenerate.
Let us discuss why we have such degeneracy in general case. 
Recall that our prescription is to determine the polynomials $\alpha_{2\ell+1,2i}(\omega)$ in 
\begin{align}
\Phi_{2\ell+1}=\frac1N\tr\,\phi^{2\ell+1}+(\nu_+-\nu_-)\sum_{i=1}^{\ell}
\alpha_{2\ell+1,2i}(\omega)\frac1N\tr\,\phi^{2i}, 
\label{mixing}
\end{align}
so that \eqref{Phi_odd_cyl} 
will be satisfied for $0\leq k\leq \ell$. 
Since $\vev{\frac1N\tr\,\phi^{2k+1}\frac1N\tr\,\phi^{2\ell+1}}_{C,0}$
contained in $\vev{\Phi_{2k+1}\Phi_{2\ell+1}}_{C,0}$ 
possibly have $\omega^{k+1}\ln\omega$, $\cdots$, $\omega^{k+\ell}\ln\omega$ terms 
which are larger than $\omega^{k+\ell+1}(\ln\omega)^2$ of our interest, 
we have to tune the polynomials $\alpha_{2\ell+1,2i}(\omega)$ 
in such a way that these $\ell$ terms all vanish for any $0\leq k\leq \ell$. 
We show that for each fixed $\ell$, 
vanishing of the leading $\ln\omega$ term ($\omega^{k+1}\ln\omega$) gives exactly the same equation 
for any $k$. Under \eqref{mixing} we have 
\begin{align}
\vev{\Phi_{2k+1}\Phi_{2\ell+1}}_{C,0}&=\vev{\frac1N\tr\,\phi^{2k+1}\frac1N\tr\,\phi^{2\ell+1}}_{C,0} \nn \\
&+(\nu_+-\nu_-)\sum_{i=1}^{\ell}\alpha_{2\ell+1,2i}(\omega)
\vev{\frac1N\tr\,\phi^{2k+1}\frac1N\tr\,\phi^{2i}}_{C,0} \nn \\
&+(\nu_+-\nu_-)\sum_{j=1}^{k}\alpha_{2k+1,2j}(\omega)
\vev{\frac1N\tr\,\phi^{2j}\frac1N\tr\,\phi^{2\ell+1}}_{C,0}
+(\text{regular terms}),
\label{cylmixing}
\end{align}
where the regular terms come from cylinder amplitudes of even-power operators. 
Let us concentrate on the leading $\ln\omega$ contribution in this equation. 
{}From (\ref{phi_odd_cyl}) and (\ref{Iexpansion}), the leading $\ln\omega$ contribution reads for $k<\ell$ 
\begin{align}
\left.\vev{\frac1N\tr\,\phi^{2k+1}\frac1N\tr\,\phi^{2\ell+1}}_{C,0}\right|_{\mbox{$\ln\omega$-leading}}
=(\nu_+-\nu_-)^2\frac{2^{k+3\ell}}{2\pi^2}\frac{(\ell-1)!}{(2\ell-1)!!}\frac{(2k+1)!!}{(k+1)!}w^{k+1}\ln\omega,
\label{phi_cyl_ln}
\end{align}
while for $k=\ell$
\begin{align}
\left.\vev{\frac1N\tr\,\phi^{2k+1}\frac1N\tr\,\phi^{2k+1}}_{C,0}\right|_{\mbox{$\ln\omega$-leading}}
=(\nu_+-\nu_-)^2\frac{2^{4k}}{\pi^2}\frac{2k+1}{k(k+1)}w^{k+1}\ln\omega.
\label{phi_cyl_ln_k=l}
\end{align}
Thus from \eqref{cyl_even_sing} 
the first and second terms on the r.h.s. in \eqref{cylmixing} have the leading $\ln\omega$ term as
$\omega^{k+1}\ln\omega$, while the third term provides $\omega^{\ell+1}\ln\omega$. 
Hence our requirement amounts to for $k<\ell$
\begin{align}
0&=\left.\vev{\frac1N\tr\,\phi^{2k+1}\frac1N\tr\,\phi^{2\ell+1}}_{C,0}\right|_{\mbox{$\ln\omega$-leading}}
\nn \\
&+(\nu_+-\nu_-)\sum_{i=1}^{\ell}\alpha^{(0)}_{2\ell+1,2i}
\left.\vev{\frac1N\tr\,\phi^{2k+1}\frac1N\tr\,\phi^{2i}}_{C,0}\right|_{\mbox{$\ln\omega$-leading}},
\label{cond1}
\end{align}
and for $k=\ell$
\begin{align}
0&=\left.\vev{\frac1N\tr\,\phi^{2\ell+1}\frac1N\tr\,\phi^{2\ell+1}}_{C,0}\right|_{\mbox{$\ln\omega$-leading}}
\nn \\
&+2(\nu_+-\nu_-)\sum_{i=1}^{\ell}\alpha^{(0)}_{2\ell+1,2i}
\left.\vev{\frac1N\tr\,\phi^{2\ell+1}\frac1N\tr\,\phi^{2i}}_{C,0}\right|_{\mbox{$\ln\omega$-leading}}.
\label{cond2}
\end{align}
Important observation here is that the ratio between two cylinder amplitudes appearing these equations 
is independent of $k$. Namely, from \eqref{cyl_even_sing} and \eqref{phi_cyl_ln} we have 
for $0\leq k<\ell$, $1\leq i\leq \ell$
\begin{align}
\frac{\left.\vev{\frac1N\tr\,\phi^{2k+1}\frac1N\tr\,\phi^{2\ell+1}}_{C,0}\right|_{\mbox{$\ln\omega$-leading}}}
{\left.\vev{\frac1N\tr\,\phi^{2k+1}\frac1N\tr\,\phi^{2i}}_{C,0}\right|_{\mbox{$\ln\omega$-leading}}}
=(\nu_+-\nu_-)\frac{2^{3\ell-i}}{\pi}\frac{(\ell-1)!}{(2\ell-1)!!}\frac{(i-1)!}{(2i-3)!!},
\label{ratio}
\end{align}
from which \eqref{cond1} becomes 
\begin{align}
\sum_{i=1}^{\ell}2^i\frac{(2i-3)!!}{(i-1)!}\alpha^{(0)}_{2\ell+1,2i}
=-\frac{2^{3\ell}}{\pi}\frac{(\ell-1)!}{(2\ell-1)!!}.
\label{generalcond}
\end{align}
The same equation is obtained for the case $k=\ell$. 
In fact, \eqref{generalcond} with $\ell=1$ reproduces $\alpha_{3,2}^{(0)}=-\frac{4}{\pi}$ 
obtained from \eqref{Phi1Phi3} and \eqref{Phi3Phi3}, 
and for $\ell=2$ it gives the same equation as \eqref{cond15_1st}. 

To summarize, we have identified 
the reason why we get the same equation for several $k$ for fixed $\ell$ 
as in \eqref{cond15_1st}. 
It originates from 
the universal structure of the leading $\ln\omega$ term that the ratio \eqref{ratio} 
does not depend on $k$. 
We expect such a universal structure persists even in higher-order $\ln\omega$ terms  
as seen in \eqref{cond15} and \eqref{cond35}. 
These observations suggest us that it would be possible to determine 
$\alpha_{2\ell+1,2i}(\omega)$ $(1\leq i\leq \ell)$ consistently 
by our requirement \eqref{Phi_odd_cyl}. 
We emphasize that dependence on $k$ and that on $\ell$ are not factorized in 
the $(\ln\omega)^2$ term \eqref{odd_cyl_ln2}, which is clearly different from the situation in 
the leading $\ln \omega$ term (\ref{phi_cyl_ln}). 

The prescription explained in this subsection is reminiscent 
of the universal and nonuniversal parts of amplitudes in matrix models for 
the two-dimensional quantum gravity. There, disk and cylinder amplitudes have regular terms with respect to 
the cosmological constant $t$, which are 
identified as contribution from surfaces constructed by 
small number of triangles in the dynamical triangulation. 
Such terms are nonuniversal, but larger than universal terms which are relevant to the continuum physics.  
Thus, in order to take the continuum limit, we have to subtract the nonuniversal parts from the 
amplitudes in advance~\cite{Fukuma:1990jw}.

\section{Three-point functions of bosons}
\label{sec:3pt_phi}
\setcounter{equation}{0}
Similar procedure to the case of the two-point functions can be used in computing three-point 
correlation functions. For the vacuum with the filling fractions $(1,0)$, 
\bea
& & \vev{\frac{1}{N}\tr\,\phi^p\, \frac{1}{N}\tr\,\phi^q\, \frac{1}{N}\tr\,\phi^r\, }^{(1,0)}_{C,0} 
= \left.\frac{\partial^2}{\partial j_p\partial j_q}\vev{\frac{1}{N}\tr\,\phi^r}^{(j)}_0\right|_{\{ j_p \}=0}
\nn \\
& & = \left.\frac{\partial^2}{\partial j_p\partial j_q}\int^1_{-1}d\zeta\,(A+B\zeta)^{\frac{r}{2}}
\tilde{\rho}(\zeta)\right|_{\{ j_p \}=0}. 
\eea
It turns out that the second-order derivatives of $A$ and $B$ with respect to $\{ j_p\}$ 
appear in total derivative terms and vanish. 
%
In terms of 
$I$'s in (\ref{I_p}), we have 
\bea
& & \hspace{-7mm}
\vev{\frac{1}{N}\tr\,\phi^p\, \frac{1}{N}\tr\,\phi^q\, \frac{1}{N}\tr\,\phi^r\, }^{(1,0)}_{C,0}  
\nn \\
& &\hspace{-3mm} = \frac{-r}{2\pi}\left.\left[\partial_{j_p}A\partial_{j_q}B\,I_{\frac{r}{2}-1}
+\frac14(\partial_{j_p}A\partial_{j_q}A+\partial_{j_p}B\partial_{j_q}B) 
\left(I_{\frac{r}{2}}-\mu^2 I_{\frac{r}{2}-1}\right)\right] \right|_{\{ j_p\}=0}\nn \\
& & +\left.\left\{\frac{p+r}{4}\partial_{j_q}B 
+ \left(\partial_{j_q}A-\frac{\mu^2}{2}\partial_{j_q}B\right)\frac{\partial}{\partial (\mu^2)}\right\} 
\right|_{\{ j_p\}=0}\vev{\frac{1}{N}\tr\,\phi^p\,\frac{1}{N}\tr\,\phi^r}^{(1,0)}_{C,0} \nn \\
& & + (p\leftrightarrow q), 
\label{three-pt_phi_10_1}
\eea
where $\left.\partial _{j_p}A\right|_{\{ j_p\}=0}$ and $\left.\partial _{j_p}B\right|_{\{ j_p\}=0}$ 
can be read off from (\ref{A_sol}) and (\ref{B_sol}). 

The amplitude for general filling fractions 
$\vev{\frac{1}{N}\tr\,\phi^p\, \frac{1}{N}\tr\,\phi^q\, \frac{1}{N}\tr\,\phi^r\, }^{(\nu_+,\nu_-)}_{C,0}$ 
is obtained by multiplying the expression (\ref{three-pt_phi_10_1}) 
by the factor $(\nu_+-\nu_-)^\sharp$ as in (\ref{nu_pm_to_10}).   

\paragraph{Case of $p=q=r=1$.}
{}From 
\bea
\left.\partial_{j_1}A\right|_{\{ j_p\}=0} & = & \frac{1}{4\sqrt{1+\omega}}\,
F\left(\frac12,\frac12,1;\frac{1}{1+\omega}\right) \nn \\
& = & -\frac{1}{4\pi}\,\ln \omega +\frac{1}{\pi}\, \ln 2+\cO(\omega\ln\omega), \\
\left.\partial_{j_1}B\right|_{\{ j_p\}=0} & = & -\frac{1}{32(1+\omega)^{3/2}}\,
F\left(\frac32,\frac32,3;\frac{1}{1+\omega}\right) \nn \\
& = & \frac{1}{4\pi}\,\ln \omega +\frac{1}{\pi}\,(1- \ln 2)+\cO(\omega\ln\omega),
\eea
the amplitude for the simplest case $p=q=r=1$ is seen to behave as 
\be
\vev{\left(\frac{1}{N}\tr\,\phi\right)^3}_{C, 0}=(\nu_+-\nu_-)^3\left[\frac{1}{16\pi^3}\,(\ln \omega)^3+ \cO((\ln \omega)^2)\right],
\label{three-pt_111}
\ee
when $\omega\sim +0$.     

\paragraph{Case of $p=q=1$ and $r=3$.}
Similarly, we see that the amplitude for $p=q=1$ and $r=3$ has the behavior 
\bea
& & \hspace{-7mm}\vev{\left(\frac{1}{N}\tr\,\phi\right)^2\,\frac{1}{N}\tr\,\phi^3}_{C,0} \nn \\
& &\hspace{-3mm} =  (\nu_+-\nu_-)^3
\left[-\frac{1}{2\pi^3}\,(\ln\omega)^2 +\frac{2}{\pi^3}\,(2\ln 2-1)\,\ln \omega \right. 
\nn \\
 & & \hspace{22mm} \left.
+\frac{8}{\pi^3}\left(-(\ln 2)^2+\ln 2 +\frac14\right) + \frac{3}{8\pi^3}\,\omega(\ln\omega)^3 
 +\cO(\omega(\ln\omega)^2)\right]. 
\label{three-pt_113} 
\eea
The terms in the first line in the r.h.s. look nonuniversal singular ones, 
and are larger than the term 
$\frac{3}{8\pi^3}\,\omega(\ln\omega)^3$ which seems universal. 
As is seen shortly, they are absorbed by the operator mixing in section~\ref{sec:mixing_phi}. 

\paragraph{Case of $p=q=1$ and $r=2$.}
We obtain the behavior of the amplitude for $p=q=1$ and $r=2$: 
\bea
\vev{\left(\frac{1}{N}\tr\,\phi\right)^2\,\frac{1}{N}\tr\,\phi^2}_{C,0} 
& = & (\nu_+-\nu_-)^2\left[-\frac{1}{8\pi^2}\,(\ln\omega)^2 
+ \frac{1}{2\pi^2}\,(2\ln 2-1)\,\ln\omega \right. \nn \\
& & \hspace{22mm} \left. +
\frac{2}{\pi^2}\left(\ln 2-(\ln 2)^2\right) + \cO(\omega(\ln \omega)^2)\right].
\label{three-pt_112}
\eea

\paragraph{Operator mixing.}
{}From (\ref{three-pt_113}) and (\ref{three-pt_112}), we see that the operator mixing (\ref{Phi}) 
discussed in section~\ref{sec:mixing_phi} 
absorbs the nonuniversal singular terms and leads to the desired result 
\be
\vev{\Phi_1^2\Phi_3}_{C,0} = (\nu_+-\nu_-)^3\left[\frac{2}{\pi^3} 
+ \frac{3}{8\pi^3}\,\omega(\ln\omega)^3  +\cO(\omega(\ln\omega)^2)\right]. 
\label{three-pt_113_mixing}   
\ee

\paragraph{Higher-point functions.}
The results obtained for the one-, two- and three-point functions of operators $\Phi_{2k+1}$ ($k=0, 1, 2, \cdots$) 
naturally suggest the form of higher-point functions as 
\be
\left.\vev{\prod_{i=1}^n\Phi_{2k_i+1}}_{C,0}\right|_{\rm sing.} = 
(\nu_+-\nu_-)^n\,(\mbox{const.})\,\omega^{2-\gamma+\sum_{i=1}^n(k_i-1)}\,(\ln \omega)^n
+(\mbox{less singular}) 
\label{higher_phi}
\ee  
with $\gamma=-1$. 
Besides the power of logarithm $(\ln \omega)^n$, it has the standard scaling behavior 
with the string susceptibility $\gamma=-1$ (the same as in the $c=-2$ topological gravity) 
and the gravitational scaling dimension $k$ of $\Phi_{2k+1}$, 
if we identify $\omega$ with ``the cosmological constant'' coupled to the 
lowest dimensional operator on 
a random surface~\cite{Knizhnik:1988ak,David:1988hj,Distler:1988jt}.

\section{Two-point functions of fermions}
\label{sec:2pt_psi}
\setcounter{equation}{0}
In this section, we consider the two-point function of fermionic operators 
\be
\vev{\frac{1}{N}\tr \,\psi^{2k+1}\frac{1}{N}\tr \,\bar{\psi}^{2k+1}} \qquad (k=0,1,2,\cdots). 
\label{2pt_psi}
\ee

It is convenient to diagonalize the matrix $\phi$ as 
$\phi = U\,{\rm diag}(\lambda_1, \cdots, \lambda_N) \,U^\dagger$ ($U\in SU(N)$) 
and to write 
\be
B=UB'U^\dagger, \qquad \psi=U\psi'U^\dagger, \qquad \bar{\psi}=U\bar{\psi}'U^\dagger. 
\ee
Then, 
\be
\vev{\psi'_{ij}\bar{\psi}'_{k\ell}}=\vev{\frac{1}{N}\frac{\delta_{i\ell}\delta_{jk}}{\lambda_i+\lambda_j}}.
\label{psi_contract}
\ee
Eq.~(\ref{2pt_psi}) can be computed by using (\ref{psi_contract}) as Wick contraction. 
The leading contribution in the large-$N$ limit comes from planar diagrams: 
\bea
& & \vev{\frac{1}{N}\tr\,\psi^{2k+1}\,\frac{1}{N}\tr\,\bar{\psi}^{2k+1}}_{C,0} \nn \\
& & = (2k+1) \frac{1}{N^{2k+1}}\sum_{i_1, i_2,\cdots, i_{2k+1}}
\vev{\frac{1}{\lambda_{i_1}+\lambda_{i_2}} \frac{1}{\lambda_{i_2}+\lambda_{i_3}}\cdots 
\frac{1}{\lambda_{i_{2k+1}}+\lambda_{i_1}}}_0  
\nn \\ 
& & = (2k+1)\int_\Omega dx_1dx_2\cdots dx_{2k+1}\,\rho(x_1)\rho(x_2)\cdots\rho(x_{2k+1})\nn \\
& & \hspace{24mm}\times {\rm P}\frac{1}{x_1+x_2}{\rm P}\frac{1}{x_2+x_3}\cdots{\rm P}\frac{1}{x_{2k+1}+x_1}. 
\label{cyl_psi}
\eea

\subsection{$\vev{\frac{1}{N}\tr\,\psi\,\frac{1}{N}\tr\,\bar{\psi}}_{C,0}$}
\label{sec:cyl_psi_k=0}
The $k=0$ case in (\ref{cyl_psi}) is given by a disk amplitude (\ref{disk_phi}) with $n=-1$:  
\bea
\vev{\frac{1}{N}\tr\,\psi\,\frac{1}{N}\tr\,\bar{\psi}}_{C,0} 
 & = & \frac12\int_\Omega dx\,\frac{1}{x}\,\rho(x) \nn \\
 & = & (\nu_+-\nu_-)\,\frac12\,(4(1+\omega))^{-1/2}\,F\left(\frac12, \frac32, 3;\frac{1}{1+\omega}\right) \nn \\
& = & (\nu_+-\nu_-)\left[\frac{4}{3\pi} +\frac{1}{\pi}\, \omega\ln \omega +\cO(\omega)\right]\qquad 
(\omega\sim +0),
\label{cyl_psi_k=0}
\eea
exhibiting $\ln\omega$ singular behavior. 
Supersymmetry invariance implies that this is equal to 
$\vev{\frac{1}{N}\tr\,(iB)\,\frac{1}{N}\tr\,\phi}_{C,0}
=\frac14\frac{\partial}{\partial\omega}\vev{\frac{1}{N}\tr\,\phi}_0$, 
interestingly which can be seen from (\ref{disk_phi_several}).  

\subsection{$\vev{\frac{1}{N}\tr\,\psi^3\,\frac{1}{N}\tr\,\bar{\psi}^3}_{C,0}$}
\label{sec:cyl_psi_k=1}
We should treat carefully the product of the principal values in (\ref{cyl_psi}) for $k\geq 1$. 
Let us compute it in the $k=1$ case: 
\be
\vev{\frac{1}{N}\tr\,\psi^3\,\frac{1}{N}\tr\,\bar{\psi}^3}_{C,0} = 
3\int_\Omega dx dy dz \,\rho(x)\rho(y)\rho(z)\,{\rm P}\frac{1}{x+y} {\rm P}\frac{1}{y+z} {\rm P}\frac{1}{z+x}. 
\label{cyl_psi_k=1_0}
\ee

We write the principal values as  
\bea
{\rm P}\frac{1}{x+y} & = & \frac12\sum_{\delta_1=\pm}
\frac{x-y+i\delta_1\epsilon}{x^2-y^2-\epsilon^2+i\delta_1\epsilon 2x}, \nn \\
{\rm P}\frac{1}{y+z} & = & \frac12\sum_{\delta_2=\pm}
\frac{y-z+i\delta_2\epsilon}{y^2-z^2-\epsilon^2+i\delta_2\epsilon 2y}, \nn \\
{\rm P}\frac{1}{z+x} & = & \frac12\sum_{\delta_3=\pm}
\frac{z-x+i\delta_3\epsilon}{z^2-x^2-\epsilon^2+i\delta_3\epsilon 2z} 
\label{prv}
\eea
with $\epsilon(>0)$ sufficiently small. 
Using 
\bea
& & \hspace{-7mm}(x-y+i\delta_1\epsilon)(y-z+i\delta_2\epsilon)(z-x+i\delta_3\epsilon) \nn \\
& & = z(x^2-y^2) +i\delta_1\epsilon (y-z)(z-x) + (\mbox{cyclic permutations}) + \cO(\epsilon^2) \nn \\
& & = z(x^2-y^2-\epsilon^2+i\delta_1\epsilon 2x) +i\delta_1\epsilon [-(y+z)x+(y-z)z] \nn \\
& & \hspace{3mm} + (\mbox{cyclic permutations}) + \cO(\epsilon^2),
\eea
where ``cyclic permutations'' means the simultaneous cyclic permutations of $(x,y,z)$ and 
$(\delta_1, \delta_2,\delta_3)$, we obtain  
\bea
& & \hspace{-7mm}{\rm P}\frac{1}{x+y}{\rm P}\frac{1}{y+z}{\rm P}\frac{1}{z+x} \nn \\
& & = \frac{1}{2^3}\sum_{\delta_1,\delta_2,\delta_3=\pm}\left[
z\,\frac{1}{y^2-z^2-\epsilon^2+i\delta_2\epsilon 2y}\,\frac{1}{z^2-x^2-\epsilon^2+i\delta_3\epsilon 2z} 
\right. \nn \\
& & \hspace{24mm}+\frac{i\delta_1\epsilon\{-(y+z)x+(y-z)z\}}{x^2-y^2-\epsilon^2+i\delta_1\epsilon 2x}\,
\frac{1}{y^2-z^2-\epsilon^2+i\delta_2\epsilon 2y}\nn \\
& & \hspace{27mm}\times\frac{1}{z^2-x^2-\epsilon^2+i\delta_3\epsilon 2z} \nn \\
& & \left. \hspace{24mm}+(\mbox{cyclic permutations}) + \cO(\epsilon^2)
\frac{}{}\right],  
\eea
and thus from \eqref{cyl_psi_k=1_0}
\bea
 & & \hspace{-7mm}\vev{\frac{1}{N}\tr \psi^3\,\frac{1}{N}\tr\bar{\psi}^3}_{C,0} \nn \\
& & =\frac{3\cdot 3}{2^3}\sum_{\delta_1,\delta_2,\delta_3=\pm} \int_\Omega dxdydz \,\rho(x)\rho(y)\rho(z)\,z 
\nn \\
& & \hspace{35mm} \times \frac{1}{y^2-z^2-\epsilon^2+i\delta_2\epsilon 2y}\,
\frac{1}{z^2-x^2-\epsilon^2+i\delta_3\epsilon 2z}  
\label{psi_psibar_k=1_careful2-1} \\
& & 
\hspace{3mm} 
+\frac{3\cdot 3}{2^3}\sum_{\delta_1,\delta_2,\delta_3=\pm} \int_\Omega dxdydz \,\rho(x)\rho(y)\rho(z)\,
\{-(y+z)x+(y-z)z\}  
\nn \\
& & \hspace{14mm} \times \frac{i\delta_1\epsilon}{x^2-y^2-\epsilon^2+i\delta_1\epsilon 2x}\,
\frac{1}{y^2-z^2-\epsilon^2+i\delta_2\epsilon 2y}\,
\frac{1}{z^2-x^2-\epsilon^2+i\delta_3\epsilon 2z}.   \nn \\
\label{psi_psibar_k=1_careful2-2} 
\eea

As $\epsilon \to 0$, the first term (\ref{psi_psibar_k=1_careful2-1}) is calculated as  
\bea
(\ref{psi_psibar_k=1_careful2-1}) & = & 
9\int_\Omega dz\,\rho(z) \,z \left(\int_\Omega dx\,\rho(x)\,{\rm P}\frac{1}{z^2-x^2}\right) 
 \left(\int_\Omega dy\,\rho(y)\,{\rm P}\frac{1}{y^2-z^2}\right) \nn \\
 & = & -\frac94\int_\Omega dz \rho(z)\,z(z^2-\mu^2)^2, 
\label{psi_psibar_k=1_careful2-3} 
\eea
by using the saddle point equation 
\be
\int_\Omega dz \rho(z){\rm P}\frac{1}{x^2-z^2}= \frac12 (x^2-\mu^2) \qquad (x\in \Omega).
\label{SPE01}
\ee 

In the second term (\ref{psi_psibar_k=1_careful2-2}), since 
\be
\sum_{\delta_1=\pm}\frac{i\delta_1\epsilon}{x^2-y^2-\epsilon^2+i\delta_3\epsilon 2x} 
\sim {\rm sgn} (x)\epsilon 2\pi\delta(x^2-y^2), 
\ee
the vicinity of $x=\pm y$ has major contribution to the $x$-integral. 
After the $x$-integral, the integrand will be proportional to 
$\epsilon \,\left({\rm P}\frac{1}{y^2-z^2}\right)^2$. 
Then the $z$-integral could have nonvanishing contribution from the neighborhood of $z=\pm y$, 
where the square of the principal value becomes singular. 

Let us evaluate (\ref{psi_psibar_k=1_careful2-2}) precisely by focusing the vicinity of 
$x=\pm y$ and $z=\pm y$ as 
\be
x=\pm y +\epsilon\tilde{x}, \qquad z=\pm y +\epsilon\tilde{z}. 
\label{poles}
\ee
At $x=y +\epsilon\tilde{x}$ and $z=y +\epsilon\tilde{z}$, the function 
\be
A\equiv \sum_{\delta_1,\delta_2,\delta_3=\pm} 
 \frac{i\delta_1\epsilon}{x^2-y^2-\epsilon^2+i\delta_1\epsilon 2x}\,
\frac{1}{y^2-z^2-\epsilon^2+i\delta_2\epsilon 2y}\,
\frac{1}{z^2-x^2-\epsilon^2+i\delta_3\epsilon 2z} 
\ee
becomes 
\be
A = \frac{1}{\epsilon^2}\frac{-1}{y^3}\,\frac{1}{\tilde{x}^2+1}\,\frac{\tilde{z}}{\tilde{z}^2+1}\,
\frac{\tilde{z}-\tilde{x}}{(\tilde{z}-\tilde{x})^2+1}.
\ee
We also get similar results for other cases in \eqref{poles}. 

Because the factor $\rho(x)\rho(y)\rho(z)\{-(y+z)x+(y-z)z\}$ in the integrand varies sufficiently 
slowly compared with $A$ around $x=\pm y$ and $z=\pm y$, 
we may fix it to its value at $x=\pm y$ and $z=\pm y$ 
in evaluating the integral. 
Then, we obtain 
\be
(\ref{psi_psibar_k=1_careful2-2}) =  \frac{3\pi^2}{4}\int_\Omega dy\,\rho(y)\frac{1}{y}
\left(\rho(y)^2-\rho(-y)^2+2\rho(y)\rho(-y)\right), 
\label{subtleterm}
\ee
after the formula 
\be
\int^\infty_{-\infty} d\tilde{x}\int^\infty_{-\infty}d\tilde{z}\,\frac{1}{\tilde{x}^2+1}\,\frac{\tilde{z}}{\tilde{z}^2+1}\,
\frac{\tilde{z}-\tilde{x}}{(\tilde{z}-\tilde{x})^2+1}
 =\frac{\pi^2}{3}
\label{sekibun} 
\ee
derived by picking up residues is used. 
Noting 
\bea
& & \rho(x)\rho(-x)=\frac{\nu_+\nu_-}{\pi^2}x^2[4-(x^2-\mu^2)^2]
=\frac{1-(\nu_+-\nu_-)^2}{4\pi^2}x^2[4-(x^2-\mu^2)^2], \nn \\
& & \rho(y)^2-\rho(-y)^2={\rm sgn}(y)\frac{\nu_+^2-\nu_-^2}{\pi^2}y^2[4-(y^2-\mu^2)^2], \nn \\
& & (\nu_+-\nu_-)\int_\Omega dy\,\rho(y)\,|y| [4-(y^2-\mu^2)^2] =\int_\Omega dy\,\rho(y)\,y [4-(y^2-\mu^2)^2] 
\label{formula_rhorho_2}
\eea
makes \eqref{subtleterm} expressed in terms of a single $\rho(y)$ as 
\bea
(\ref{psi_psibar_k=1_careful2-2}) & = & \frac98\int_\Omega dy\,\rho(y)\,y[4-(y^2-\mu^2)^2] \nn \\
& & -\frac38(\nu_+-\nu_-)^2\int_\Omega dy\,\rho(y)\,y [4-(y^2-\mu^2)^2]. 
\label{psi_psibar_k=1_careful2-4}
\eea

Thus, the final result is 
\bea
\vev{\frac{1}{N}\tr \psi^3\,\frac{1}{N}\tr\bar{\psi}^3}_{C,0} & = & 
(\ref{psi_psibar_k=1_careful2-3}) + (\ref{psi_psibar_k=1_careful2-4}) \nn \\
 & = & \frac98\vev{\frac{1}{N}\tr\left[4\phi-3\phi(\phi^2-\mu^2)^2\right]}_0 \nn \\
 & & -\frac38(\nu_+-\nu_-)^2\vev{\frac{1}{N}\tr\left[4\phi-\phi(\phi^2-\mu^2)^2\right]}_0.  
\label{cyl_psi_k=1_careful1-f}
\eea 
It is a linear combination of the planar expectation values of odd-power operators 
computed in section~\ref{sec:1pt_phi}. 
For $\omega\sim +0$, (\ref{cyl_psi_k=1_careful1-f}) behaves as 
\bea
\vev{\frac{1}{N}\tr \,\psi^3\,\frac{1}{N}\tr\,\bar{\psi}^3}_{C,0} 
& = &(\nu_+-\nu_-)\left[\frac{192}{35\pi}+\frac{48}{35\pi}\,\omega-\frac{18}{\pi}\,\omega^2\ln\omega 
-\frac{9}{5\pi}(49-40\ln 2)\,\omega^2 \right. \nn \\
& & \hspace{21mm}\left. -\frac{117}{2\pi}\,\omega^3\ln\omega +\cO(\omega^3)\right]
\nn \\
& & 
-(\nu_+-\nu_-)^3\left[\frac{512}{105\pi}+\frac{192}{35\pi}\,\omega-\frac{24}{5\pi}\,\omega^2
-\frac{6}{\pi}\,\omega^3\ln\omega +\cO(\omega^3)\right] .\nn \\
& & 
\label{cyl_psi_k=1}
\eea


We can check that this result can be obtained by a different treatment of the principle value 
from \eqref{prv} as 
\bea
& & {\rm P}\frac{1}{x+y} = \frac12\sum_{\delta_1=\pm}\frac{1}{x+y+i\delta_1\epsilon}, \qquad 
{\rm P}\frac{1}{y+z} = \frac12\sum_{\delta_2=\pm}\frac{1}{y+z+i\delta_2\epsilon}, \nn \\
& & {\rm P}\frac{1}{z+x} = 
\frac12\sum_{\delta_3=\pm}\frac{z-x+i\delta_3\epsilon}{z^2-x^2-\epsilon^2+i\delta_3\epsilon 2z}
\eea
This strongly supports the validity of the computations here.

\subsection{Operator mixing}
\label{sec:mixing_psi}
In the derivation of (\ref{cyl_psi_k=1}), the factor $(\nu_+-\nu_-)^3$ appears in a nontrivial way from 
the product of three $\rho$'s 
in (\ref{subtleterm}). 
For general $k$, it is expected that terms proportional to $(\nu_+-\nu_-)^{2k+1}$ arise in 
$\vev{\frac{1}{N}\tr \,\psi^{2k+1}\,\frac{1}{N}\tr\,\bar{\psi}^{2k+1}}_{C,0}$ from 
the product of $(2k+1)$ $\rho$'s. Hence terms proportional to $(\nu_+-\nu_-)^{2k+1}$ are considered 
to be characteristic of 
the amplitude $\vev{\frac{1}{N}\tr \,\psi^{2k+1}\,\frac{1}{N}\tr\,\bar{\psi}^{2k+1}}_{C,0}$.  

In (\ref{cyl_psi_k=1}), the leading singular term for $\omega\sim +0$ 
is not $\frac{6}{\pi}\,(\nu_+-\nu_-)^3\omega^3\ln\omega$, but 
$-\frac{18}{\pi}\,(\nu_+-\nu_-)\omega^2\ln\omega$. 
Similarly to what was discussed for the operators $\Phi_{2k+1}$ in section~\ref{sec:mixing_phi}, 
we can take a new basis of operators to make the $(\nu_+-\nu_-)^3\omega^3\ln\omega$ term dominant 
in the singular terms. 
Let us take the new basis as  
\bea
& \Psi_1\equiv \frac{1}{N}\tr\,\psi, \qquad & \bar{\Psi}_1\equiv \frac{1}{N}\tr\,\bar{\psi}, \nn \\
&  \Psi_3\equiv \frac{1}{N}\tr\,\psi^3 +(\mbox{mixing}), \qquad & 
\bar{\Psi}_3\equiv \frac{1}{N}\tr\,\bar{\psi}^3+(\mbox{mixing}), 
\label{mixing_psi_1}
\eea
where ``mixing'' means operators to be added 
so that 
\be
\left.\vev{\Psi_{2k+1}\bar{\Psi}_{2\ell+1}}_{C,0}\right|_{\rm sing.} 
= \delta_{k,\ell}\,v_k\,(\nu_+-\nu_-)^{2k+1}\omega^{2k+1}\ln\omega 
+(\mbox{less singular})
\label{mixing_psi_2}
\ee
with $v_k$ constants holds for $k, \ell=0,1$. 

As ``mixing'' operators in $\Psi_3$, we consider fermionic operators 
which have lower powers of $\psi$ than $\psi^3$ and preserve the $Q$-exactness: 
$\Psi_3=Q\Xi$ as $\frac1N\tr\,\psi^3=Q\left(\frac1N\tr\,\phi\psi^2\right)$. 
They are allowed to be multiplied by polynomials of $\omega$ 
as $\alpha_{2\ell+1,2i}(\omega)$ in \eqref{mixing}. 
We find $\frac{1}{N}\tr\,\{(iB-\phi^2+\mu^2)\psi\}$ as an interesting candidate by noting that 
\bea
 & & \hspace{-7mm} 
\vev{\frac{1}{N}\tr\,\{(iB-\phi^2+\mu^2)\psi\}\,\frac{1}{N}\tr\,\{(iB-\phi^2+\mu^2)\bar{\psi}\}}_{C,0} \nn\\
& & = -\vev{\frac{1}{N}\tr\,(\psi\bar{\psi})}_0 = -\int_\Omega dxdy\,\rho(x)\rho(y)\,{\rm P}\frac{1}{x+y} 
 = -\int_\Omega dx\,\rho(x)x(x^2-\mu^2) \nn \\
& & = (\nu_+-\nu_-)\left[-\frac{128}{105\pi}+\frac{32}{15\pi}\,\omega +\frac{4}{\pi}\,\omega^2\ln\omega 
+\frac{2}{3\pi}(23-24\ln 2)\,\omega^2 + \frac{5}{\pi}\,\omega^3\ln\omega \right. \nn \\
& & \hspace{24mm} \left.+ \frac{1}{6\pi}(67-120\ln 2)\,\omega^3 
+ \cO(\omega^4\ln\omega)\frac{}{}\right] \qquad (\omega\sim +0),
\eea
where (\ref{psi_contract}) and the saddle point equation were utilized in the second line, and that 
\be
\vev{\frac{1}{N}\tr\,\{(iB-\phi^2+\mu^2)\psi\}\,\frac{1}{N}\tr\,\bar{\psi}^{2k+1}}_{C,0}
= \vev{\frac{1}{N}\tr\,\psi^{2k+1}\,\frac{1}{N}\tr\,\{(iB-\phi^2+\mu^2)\bar{\psi}\}}_{C,0}=0 
\ee
with $k=0,1$.

It can be shown from these equations that either of  
\bea
\Psi_3 & = & \frac{1}{N}\tr\,\psi^3
+\frac{3}{\sqrt{2}}\,(1+\omega+\cO(\omega^2))\frac{1}{N}\tr\,\{(iB-\phi^2+\mu^2)\psi\}, \nn \\
\bar{\Psi}_3 & = & \frac{1}{N}\tr\,\bar{\psi}^3
+\frac{3}{\sqrt{2}}\,(1+\omega+\cO(\omega^2))\frac{1}{N}\tr\,\{(iB-\phi^2+\mu^2)\bar{\psi}\} 
\eea
and 
\bea
\Psi_3 & = & \frac{1}{N}\tr\,\psi^3
-\frac{3}{\sqrt{2}}\,(1+\omega+\cO(\omega^2))\frac{1}{N}\tr\,\{(iB-\phi^2+\mu^2)\psi\}, \nn \\
\bar{\Psi}_3 & = & \frac{1}{N}\tr\,\bar{\psi}^3
-\frac{3}{\sqrt{2}}\,(1+\omega+\cO(\omega^2))\frac{1}{N}\tr\,\{(iB-\phi^2+\mu^2)\bar{\psi}\}
\eea
does the job (\ref{mixing_psi_2}) with $v_0=\frac{1}{\pi}$ and $v_1=\frac{6}{\pi}$, 
where we have implicitly assumed the same coefficient in the mixing for $\Psi_3$ and $\bar{\Psi}_3$.  

Eq.~(\ref{mixing_psi_2}) indicates that $\Psi_{2k+1}$ and $\bar{\Psi}_{2k+1}$ have the gravitational scaling 
dimension $k$ (the same as $\Phi_{2k+1}$), besides the $\ln\omega$ correction.

\section{Correspondence to type IIA superstring theory}
\label{sec:discussion}
\setcounter{equation}{0}

We computed various amplitudes of the supersymmetric double-well matrix model (\ref{S}), and 
observed critical behavior with powers of logarithm in planar correlation functions 
among operators (\ref{phi_psi_odd}),  
which are not belonging to the observables in the $c=-2$ topological gravity. 
The result reminds us of the logarithmic scaling violation of 
the two-dimensional string (two-dimensional quantum gravity coupled to the $c=1$ matter). 
Note that the Penner model is known \cite{Distler:1990mt} as a zero-dimensional matrix model 
to describe the two-dimensional bosonic string whose target space 
is ${\bf R}\times S^1$, where ${\bf R}$ is the Liouville direction, and 
$S^1$ is the matter direction compactified to the self-dual 
radius. 
Thus it is interesting to proceed with 
an interpretation of our matrix model as a supersymmetric version of the Penner model 
describing two-dimensional string theory with target-space supersymmetry. 

Let us begin with symmetry consideration. 
Suppose $\psi$ and $\bar{\psi}$ are regarded as target-space fermions in the corresponding 
superstring theory, i.e. 
$\psi$ is interpreted as an operator in the (NS, R) sector and $\bar{\psi}$ in the (R, NS) sector in the RNS formalism. 
Then, under the so-called $(-1)^{{\bf F}_L}$ and $(-1)^{{\bf F}_R}$ transformations changing the signs of 
operators in the left-moving Ramond sector and those in the right-moving 
Ramond sector respectively, they transform as 
\bea
(-1)^{{\bf F}_L}: & & \psi \to \psi, \qquad \bar{\psi}\to -\bar{\psi}, \label{FL}\\ 
(-1)^{{\bf F}_R}: & & \psi \to -\psi, \qquad \bar{\psi}\to \bar{\psi}. 
\eea
In order for the matrix model action (\ref{S}) to be invariant under the transformations, $B$ and $\phi$ 
should transform as 
\bea
(-1)^{{\bf F}_L}: & & B \to B, \qquad \phi\to -\phi, \\ 
(-1)^{{\bf F}_R}: & & B \to B, \qquad \phi\to -\phi. \label{FR}
\eea
This indicates that $B$ corresponds to an operator in the (NS, NS) sector, and $\phi$ in the (R, R) sector. 
In this interpretation, the Ramond sector is considered to be responsible for 
the logarithmic singular behavior seen in correlators 
among operators (\ref{phi_psi_odd}).  

In fact, two-dimensional type II superstring theory with the identical target space, 
$(\varphi, x)\in (\mbox{Liouville direction}, \,\mbox{$S^1$ with the self-dual radius})$, is constructed 
in~\cite{Kutasov:1990ua,Murthy:2003es,Ita:2005ne,Grassi:2005kc}. 
The (holomorphic) energy-momentum tensor on the string world-sheet 
except ghosts' part is 
\be
T_m = -\frac12 (\partial x)^2 -\frac12 \psi_x\partial\psi_x -\frac12(\partial \varphi)^2 
+\frac{Q}{2}\partial^2\varphi -\frac12\psi_\ell\partial \psi_\ell
\ee
with $Q=2$. $\psi_x$ and $\psi_\ell$ are superpartners of $x$ and $\varphi$, respectively. 
Target-space supercurrents in the type IIA theory  
\be
q_+(z)=e^{-\frac12\phi(z)-\frac{i}{2}H(z)-ix(z)}, \qquad 
\bar{q}_-(\bar{z})= e^{-\frac12\bar{\phi}(\bar{z})+\frac{i}{2}\bar{H}(\bar{z})+i\bar{x}(\bar{z})}
\ee
can exist only for the $S^1$ target space of the self-dual radius, which comes 
from the GSO projection implemented by imposing locality of the supercurrents 
with operators of the theory. 
$\phi$ ($\bar{\phi}$) is the holomorphic (anti-holomorphic) bosonized superconformal 
ghost, and the fermions are bosonized as 
\be
\psi_\ell \pm i\psi_x=\sqrt{2}\,e^{\mp iH}, \qquad 
\bar{\psi}_\ell \pm i \bar{\psi}_x= \sqrt{2}\,e^{\mp i\bar{H}}.
\ee 
Then the supercharges 
\be
Q_+=\oint \frac{dz}{2\pi i}\,q_+(z), \qquad \bar{Q}_- = \oint \frac{d\bar{z}}{2\pi i}\,\bar{q}_-(\bar{z})
\ee
are nilpotent $Q_+^2=\bar{Q}_-^2=0$, which indeed matches the supercharges $Q$ and $\bar{Q}$ 
in the matrix model in \eqref{QSUSY} and \eqref{QbarSUSY}.   

The spectrum except special massive states is represented by the NS ``tachyon'' vertex operator 
(in $(-1)$ picture): 
\be
T_k = e^{-\phi+ikx+p_\ell\varphi}, \qquad 
\bar{T}_{\bar{k}} = e^{-\bar{\phi}+i\bar{k}\bar{x}+p_\ell\bar{\varphi}}, 
\ee
and by the R vertex operator (in $(-\frac12)$ picture): 
\be
V_{k, \,\epsilon}=e^{-\frac12\phi+\frac{i}{2}\epsilon H+ikx +p_\ell\varphi}, \qquad 
\bar{V}_{\bar{k},\,\bar{\epsilon}}= 
e^{-\frac12\bar{\phi}+\frac{i}{2}\bar{\epsilon} \bar{H}+i\bar{k}\bar{x} +p_\ell\bar{\varphi}} 
\ee
with $\epsilon, \bar{\epsilon}=\pm 1$. 
Locality with the supercurrents, mutual locality, superconformal invariance (including the Dirac equation 
constraint) and the level matching condition 
determine physical vertex operators. As discussed in~\cite{Ita:2005ne}, there are two consistent sets of 
physical vertex operators - ``momentum background'' and ``winding background''. 
Let us consider the ``winding background''~\footnote{
We can repeat the parallel argument for ``momentum background'' in the type IIB theory, which 
is equivalent to the ``winding background'' in the type IIA theory through T-duality with respect to 
the $S^1$ direction. 
}. 
The physical spectrum is given by 
\bea
\mbox{(NS, NS)}: \quad &  T_k\,\bar{T}_{-k}  & \quad (k\in {\bf Z}+\frac12), \nn \\
\mbox{(R$+$, R$-$)}: \quad &  
V_{k, \,+1}\,\bar{V}_{-k, \,-1} & \quad (k=\frac12,\,\frac32, \cdots), \nn \\
\mbox{(R$-$, R$+$)}: \quad &  
V_{-k, \,-1}\,\bar{V}_{k,\, +1} & \quad(k=0,1,2,\cdots), \nn \\
 \mbox{(NS, R$-$)}: \quad &  
T_{-k}\,\bar{V}_{-k,\,-1}  & \quad (k=\frac12,\,\frac32, \cdots), \nn \\
\mbox{(R$+$, NS)}: \quad &  
V_{k, \,+1}\, \bar{T}_k\,  & \quad (k=\frac12,\,\frac32, \cdots), 
\eea  
where we take a branch of $p_\ell=1-|k|$ satisfying the locality bound~\cite{Seiberg:1990eb}. 
We can see that the vertex operators 
\be
V_{\frac12, \,+1}\,\bar{V}_{-\frac12, \,-1}, \qquad T_{-\frac12}\,\bar{V}_{-\frac12, \,-1}, \qquad 
V_{\frac12,\,+1}\,\bar{T}_{\frac12}, \qquad  T_{-\frac12}\,\bar{T}_{\frac12}
\ee 
form a quartet under $Q_+$ and $\bar{Q}_-$:~\footnote{
We here assume that $Q_+$ commutes with $\bar{T}_{\bar{k}}$ and 
anti-commutes with $\bar{V}_{\bar{k},\,\bar{\epsilon}}$, 
and that $\bar{Q}_-$ commutes with $T_k$ and anti-commutes with 
$V_{k,\,\epsilon}$. It is plausible from the statistics in the target space. 
In ref.~\cite{kuroki-sugino}, we 
introduce cocycle factors to the vertex operators in order to realize the (anti-)commutation properties. 
}  
\bea
& 
[Q_+,  V_{\frac12, \,+1}\,\bar{V}_{-\frac12, \,-1} ]  = T_{-\frac12}\,\bar{V}_{-\frac12, \,-1}, 
& \{ Q_+, T_{-\frac12}\,\bar{V}_{-\frac12, \,-1}\}=0, 
\nn \\
 & 
\{ Q_+, V_{\frac12, \,+1}\, \bar{T}_{\frac12}\} = T_{-\frac12}\,\bar{T}_{\frac12}, 
& [Q_+, T_{-\frac12}\,\bar{T}_{\frac12}]=0, 
\label{quartet_+} 
\\
& 
[\bar{Q}_-, V_{\frac12,\,+1}\,\bar{V}_{-\frac12, \,-1}] = -V_{\frac12,\,+1}\, \bar{T}_{\frac12}, 
& \{ \bar{Q}_-, V_{\frac12,\,+1}\, \bar{T}_{\frac12}\}= 0, 
\nn \\
 & 
\{ \bar{Q}_-, T_{-\frac12}\,\bar{V}_{-\frac12,\,-1}\} = T_{-\frac12}\,\bar{T}_{\frac12}, 
 & [\bar{Q}_-, T_{-\frac12}\,\bar{T}_{\frac12}]=0. 
\label{quartet_-}
\eea

Notice that (\ref{quartet_+}) and (\ref{quartet_-}) are isomorphic to (\ref{QSUSY}) and (\ref{QbarSUSY}), 
respectively. 
It leads to correspondence of single-trace operators in the matrix model to integrated 
vertex operators in the type IIA theory:  
\bea
\Phi_1= \frac{1}{N}\tr\,\phi & \Leftrightarrow &  
\int d^2z\, V_{\frac12, \,+1}(z)\,\bar{V}_{-\frac12, \,-1}(\bar{z}) , \nn \\
\Psi_1 = \frac{1}{N}\tr\,\psi & \Leftrightarrow &  
\int d^2z\, T_{-\frac12}(z)\,\bar{V}_{-\frac12, \,-1}(\bar{z}), \nn \\
\bar{\Psi}_1=\frac{1}{N}\tr\,\bar{\psi} & \Leftrightarrow &  
\int d^2z\, V_{\frac12, \,+1}(z)\, \bar{T}_{\frac12}(\bar{z}), \nn \\
\frac{1}{N}\tr\,(-iB) & \Leftrightarrow &  
\int d^2z\, T_{-\frac12}(z)\,\bar{T}_{\frac12}(\bar{z}),  
\label{mat_IIA_op1}
\eea
which is consistent with the identification in (\ref{FL})--(\ref{FR}). 
Furthermore, it is natural to extend (\ref{mat_IIA_op1}) to case of higher $k(=1,2,\cdots)$ as 
\bea
\Phi_{2k+1}=\frac{1}{N}\tr\,\phi^{2k+1}+\mbox{(mixing)} & \Leftrightarrow & 
\int d^2z\, V_{k+\frac12, \,+1}(z)\,\bar{V}_{-k-\frac12, \,-1}(\bar{z}) , \nn \\
\Psi_{2k+1}=\frac{1}{N}\tr\,\psi^{2k+1}+\mbox{(mixing)} & \Leftrightarrow &  
\int d^2z\, T_{-k-\frac12}(z)\,\bar{V}_{-k-\frac12, \,-1}(\bar{z}), \nn \\
\bar{\Psi}_{2k+1} = \frac{1}{N}\tr\,\bar{\psi}^{2k+1}+\mbox{(mixing)} & \Leftrightarrow &  
\int d^2z\, V_{k+\frac12, \,+1}(z)\, \bar{T}_{k+\frac12}(\bar{z}). 
\label{mat_IIA_op2}
\eea
Since the ``tachyons'' of the negative winding 
$\int d^2z\, T_{-k-\frac12}(z)\,\bar{T}_{k+\frac12}(\bar{z})$ ($k=0,1,2,\cdots$) are invariant 
under $Q_+$ and $\bar{Q}_-$, 
they are expected to be mapped to $\{\frac{1}{N}\tr\,(-iB)^{k+1}\}$ 
($k=0,1,2,\cdots$) perhaps with some mixing terms. 
We see in (\ref{mat_IIA_op2}) that the powers of matrices are interpreted as windings or momenta in the 
$S^1$ direction of the type IIA theory. 
Although such interpretation is not usual in matrix models for two-dimensional quantum gravity coupled to 
$c<1$ matters, refs.~\cite{Imbimbo:1995yv,Mukhi:2003sz} show that a positive power $k$ 
of a matrix variable in the Penner model correctly describe the ``tachyons'' with 
negative momentum $-k$ in the $c=1$ string on $S^1$, 
which is in harmony with our interpretation. 
In \cite{Imbimbo:1995yv,Mukhi:2003sz}, the positive momentum ``tachyons''   
are represented by introducing source terms of an external matrix via the Kontsevich-Miwa transformation 
in the Penner model. 
In turn, it is natural to expect in our case that the positive winding ``tachyons'' 
$\int d^2z\, T_{-k-\frac12}(z)\,\bar{T}_{k+\frac12}(\bar{z})$ ($k=-1,-2,\cdots$)
in the type IIA theory are expressed 
in a similar manner in the matrix model. 

Thus, from the argument here based on symmetry properties and spectrum of the matrix model and 
the type IIA theory, 
we have a plausible reason to expect correspondence between them. 
The vertex operators in the (R$-$, R$+$) sector are singlets under the target-space supersymmetries 
$Q_+$ and $\bar{Q}_-$, and seem to have no counterparts in the matrix model. 
As we have seen in sections~\ref{sec:1pt_phi}, \ref{sec:3pt_phi} and \ref{sec:2pt_psi}, 
one- and three-point functions of $\Phi_{2k+1}$ and two-point functions of 
fermions $\Psi_{2k+1}$, $\bar{\Psi}_{2k+1}$ are nonvanishing.  
From the point of view of our correspondence \eqref{mat_IIA_op2}, 
it implies that correlators of operators with nonzero Ramond charges do not vanish, 
and hence the matrix model should 
correspond to the type IIA theory on a background of (R$-$, R$+$) operators. 
Furthermore (\ref{Bn_torus}) suggests that the torus free energy of the matrix model $F_1$ 
in (\ref{F}) vanishes. 
Although it is different from the results of calculations in the string theory 
on the trivial background~\cite{Kutasov:1990ua,Murthy:2003es}, 
we can find a consistent (R$-$, R$+$) background in the type IIA theory which gives vanishing 
torus partition function in agreement with the result of the matrix model. 
Details will be presented in ref.~\cite{kuroki-sugino}. 

So far we have seen the correspondence between our matrix model \eqref{S} 
and the two-dimensional type IIA theory with an RR background 
at the level of symmetries and spectrum. 
This correspondence is in fact more than that and quantitative. 
In the next paper~\cite{kuroki-sugino}, we will calculate various amplitudes 
in the type IIA theory and find surprisingly that they also exhibit powers of logarithm 
as in the amplitudes in the matrix model presented in this paper.   
Thus we will directly see the correspondence quantitatively between amplitudes in both sides 
in \eqref{mat_IIA_op2}. 

Our interpretation of the matrix model as superstrings does not rely on the picture of unstable D-branes 
discussed in~\cite{McGreevy:2003kb}-\cite{McGreevy:2003dn}. 
It will be interesting to consider a relation 
of the matrix model with D-branes. 
In fact, it is quite likely that our matrix model would be interpreted as an effective action 
of D-branes and that the correspondence to the IIA theory could be a consequence 
of the open-closed duality.  
For the aim, ref.~\cite{Mukherjee:2005aq} would be helpful, 
which discusses FZZT branes \cite{Fateev:2000ik,Teschner:2000md} 
in the Kontsevich-Penner model.

\section*{Acknowledgements}
We would like to thank Yasuaki~Hikida, Kazuo~Hosomichi, Hirotaka~Irie, Satoshi~Iso, 
Shoichi~Kawamoto, Sanefumi~Moriyama, Koichi~Murakami, Shinsuke~Nishigaki, Kazumi Okuyama, Hidehiko~Shimada, 
Yuji~Sugawara, Hiroshi~Suzuki, Tadashi Takayanagi and Naoto~Yokoi for useful discussions and comments. 
F.~S. thanks KITP Santa Barbara for hospitality during his stay (February, 2012), 
where a part of this work was done. 
The authors thank Osaka City University and the Yukawa Institute for Theoretical Physics at Kyoto University. 
Discussions during the conference ``Progress in Quantum Field Theory and String Theory'' (April, 2012) 
and the YITP workshop ``Field Theory and String Theory'' (YITP-W-12-05, July, 2012) 
were useful to complete this work. 
The work of T.~K. is supported in part by Rikkyo University Special Fund for Research, 
and the work of F.~S. is supported in part by a Grant-in-Aid for Scientific Research (C), 21540290.

\appendix
\section{Analysis via Schwinger-Dyson equations}
\label{app:SD}
\setcounter{equation}{0}
In this appendix we analyze correlation functions of even-power operators of $\phi$ 
via Schwinger-Dyson equations.  
For this purpose let us consider the Schwinger-Dyson equations 
\bea
0 & = & \sum_{k=0}^m\vev{\frac{1}{N}\tr\,\phi^k\,\frac{1}{N}\tr\,\phi^{m-k}}
-2i\vev{\frac{1}{N}\tr(B\phi^{m+2})} \nn \\
  & & -\vev{\frac{1}{N}\tr\left(\phi^{m+1}(\bar{\psi}\psi-\psi\bar{\psi})\right)}, 
\label{SD_1}\\
0 & = & \frac{n}{N^2}\vev{\frac{1}{N}\tr\,\phi^{n+m}} 
+\sum_{k=0}^m\vev{\frac{1}{N}\tr\,\phi^n\,\frac{1}{N}\tr\,\phi^k\,\frac{1}{N}\tr\,\phi^{m-k}}  \nn \\
& & -2i\vev{\frac{1}{N}\tr\,\phi^n\,\frac{1}{N}\tr(B\phi^{m+2})}
-\vev{\frac{1}{N}\tr\,\phi^n\,\frac{1}{N}\tr\left(\phi^{m+1}(\bar{\psi}\psi-\psi\bar{\psi})\right)}, \nn \\
& & \label{SD_2}
\eea
which are derived from 
\bea
0 & = & \int d^{N^2}B \, d^{N^2}\phi \, d^{N^2}\psi \, d^{N^2}\bar{\psi} \,
\sum_{\alpha=1}^{N^2} \frac{\partial}{\partial\phi^\alpha}\left[\tr(\phi^{m+1}t^\alpha)\,e^{-S}\right], 
\\
0 & = & \int d^{N^2}B \, d^{N^2}\phi \, d^{N^2}\psi \, d^{N^2}\bar{\psi} \,
\sum_{\alpha=1}^{N^2} \frac{\partial}{\partial\phi^\alpha}\left[\tr(\phi^n)\,\tr(\phi^{m+1}t^\alpha)\,e^{-S}\right], 
\eea
respectively. 
Here, $\phi$ is expanded by a basis of $N\times N$ hermitian matrices $\{t^\alpha\}$ ($\alpha=1, \cdots, N^2$) 
as 
$\phi=\sum_{\alpha=1}^{N^2}\phi^\alpha t^\alpha$. 

By diagonalizing $\phi$ and integrating out fermions, 
$\frac{1}{N}\tr\left(\phi^{m+1}(\bar{\psi}\psi-\psi\bar{\psi})\right)$ in (\ref{SD_1}) and (\ref{SD_2})  
becomes  
$-\frac{1}{N^2}\sum_{i,j=1}^N\frac{\lambda_i^{m+1}+\lambda_j^{m+1}}{\lambda_i+\lambda_j}$, 
where $\lambda_i$ ($i=1, \cdots, N$) are the eigenvalues of $\phi$. 
Note that when $m$ is even, this is further reduced to the polynomial: 
\be
\frac{\lambda_i^{m+1}+\lambda_j^{m+1}}{\lambda_i+\lambda_j}= \sum_{k=0}^m(-1)^k\lambda_i^k\lambda_j^{m-k}. 
\ee
Thus, for $m=2\ell$ ($\ell=0,1,2,\cdots$), (\ref{SD_1}) and (\ref{SD_2}) can be rewritten as   
\be
\sum_{k=0}^\ell\vev{\frac{1}{N}\tr\,\phi^{2k}\,\frac{1}{N}\tr\,\phi^{2(\ell-k)}} 
= i\vev{\frac{1}{N}\tr(B\phi^{2\ell+2})},
\label{SD_3}
\ee
and 
\bea
& & \frac{1}{N^2}\frac{n}{2}\vev{\frac{1}{N}\tr\,\phi^{n+2\ell}} + 
\sum_{k=0}^\ell\vev{\frac{1}{N}\tr\,\phi^n\,\frac{1}{N}\tr\,\phi^{2k}\,\frac{1}{N}\tr\,\phi^{2(\ell-k)}}  \nn \\
& & = i\vev{\frac{1}{N}\tr\,\phi^n\,\frac{1}{N}\tr(B\phi^{2\ell+2})},
\label{SD_4}
\eea
respectively. 

In terms of the resolvent for $\phi^2$: 
\be
R_2(z) \equiv \frac{1}{N}\tr\,\frac{1}{z-\phi^2}, 
\ee
eq.~(\ref{SD_3}) is expressed as 
\be
z\vev{R_2(z)^2} = (z^2-\mu^2z) \vev{R_2(z)} -z+\mu^2 -C, \qquad 
C\equiv \vev{\frac{1}{N}\tr\,\phi^2}. 
\label{SD_5}
\ee
Also, (\ref{SD_4}) for $n$ even is as 
\bea
& & \frac{1}{N^2}\partial_w\left(wD_{z'}(z,w)\vev{R_2(z')}\right) + z\vev{R_2(z)^2\,R_2(w)} \nn \\
& & =(z^2-\mu^2z)\vev{R_2(z)\,R_2(w)} +(-z+\mu^2)\vev{R_2(w)} -\vev{\frac{1}{N}\tr\,\phi^2\,R_2(w)} \nn \\
& & 
\label{SD_6}
\eea
with 
\be
D_{z'}(z,w)f(z') \equiv \frac{f(z)-f(w)}{z-w}.
\ee

\subsection{Solution of (\ref{SD_5}) and (\ref{SD_6})}
Eqs.~(\ref{SD_5}), (\ref{SD_6}) can be solved by the genus expansion. 
By writing the genus expansion of $C$ as in \eqref{genus_exp}: 
\be
C = C_0 + \frac{1}{N^2}C_1 + \frac{1}{N^4}C_2 +\cdots, \qquad C_h\equiv \vev{\frac{1}{N}\tr\,\phi^2}_h, 
\ee
the $\cO(N^0)$ part of (\ref{SD_5}) reads  
\be
z\left(\vev{R_2(z)}_0\right)^2 = (z^2-\mu^2z)\vev{R_2(z)}_0 -z+\mu^2-C_0.
\label{R2(z)eq}
\ee
Its solution that does not blow up as $z\to \infty$ is 
\be
\vev{R_2(z)}_0 = \frac12\left[z-\mu^2-\sqrt{(z-\mu^2)^2-4+\frac{4(\mu^2-C_0)}{z}}\right]. 
\ee
Assuming that $\vev{R_2(z)}_0$ should be analytic except one cut 
on the positive real axis $[a^2, b^2]$ ($0<a^2<b^2$) 
determines $C_0$ as 
\be
C_0 = \mu^2. 
\ee
Thus, we obtain 
\bea
& & \vev{R_2(z)}_0 = \frac12\left[z-\mu^2-\sqrt{(z-a^2)(z-b^2)}\right], 
\label{R2_0} \\
& & a^2 = -2+\mu^2, \qquad b^2= 2+\mu^2, \nn 
\eea
which coincides with the result from the saddle point equation. 
Note that it holds for general filling fractions $(\nu_+,\nu_-)$. 

{}From the $\cO(N^{-2})$ part of (\ref{SD_5}): 
\be
\vev{R_2(z)^2}_{C,0} -\sqrt{(z-a^2)(z-b^2)}\vev{R_2(z)}_1 = -\frac{C_1}{z}
\label{SD_7}
\ee
and the $\cO(N^{-2})$ part of (\ref{SD_6}): 
\bea
& & \partial_w\left(wD_{z'}(z,w)\vev{R_2(z')}_0\right) + z\vev{R_2(w)}_0 \vev{R_2(z)^2}_{C, 0} \nn \\
& & -z\sqrt{(z-a^2)(z-b^2)}\left[\vev{R_2(z)\,R_2(w)}_{C, 0} +\vev{R_2(w)}_0\vev{R_2(z)}_1\right] \nn \\
& & = -\vev{\frac{1}{N}\tr\,\phi^2\,R_2(w)}_{C, 0} -C_1\vev{R_2(w)}_0 
\eea
for both of which we have used (\ref{R2_0}), 
we obtain 
\bea
 & & \partial_w\left(wD_{z'}(z,w)\vev{R_2(z')}_0\right)-z\sqrt{(z-a^2)(z-b^2)}\vev{R_2(z)\,R_2(w)}_{C, 0}
\nn \\
& & = -\vev{\frac{1}{N}\tr\,\phi^2\,R_2(w)}_{C, 0}. 
\label{SD_8}
\eea
Applying 
\be
\frac{\partial}{\partial(\mu^2)}\vev{R_2(w)}_0 = \vev{\frac{1}{N}\tr\,\phi^2\,R_2(w)}_{C, 0}   
\ee
and (\ref{R2_0}) to (\ref{SD_8}) leads to 
\be
\vev{R_2(z)\,R_2(w)}_{C, 0} = \frac14\frac{1}{(z-w)^2}\left[\sqrt{\frac{(z-a^2)(w-b^2)}{(z-b^2)(w-a^2)}} 
+\sqrt{\frac{(z-b^2)(w-a^2)}{(z-a^2)(w-b^2)}}-2\right]. 
\label{evencyl}
\ee
It is easy to confirm that this equation yields \eqref{cyl1_nu_pm} for $p$, $q$ even. 

In particular, 
\be
\vev{R_2(z)^2}_{C,0}=\frac{1}{(z-a^2)^2(z-b^2)^2},
\ee
and requiring the regularity of $\vev{R_2(z)}_1$ at $z=0$ in \eqref{SD_7} gives 
\bea
 & & \vev{R_2(z)}_1 = \frac{1}{(z-a^2)^{5/2}(z-b^2)^{5/2}}, 
\label{R2_1} \\
 & & C_1 = 0. \label{C1}
\eea 
Eq.~(\ref{C1}) means that the torus free energy is a constant independent of $\mu^2$. 
We explicitly computed $\vev{\frac{1}{N}\tr\,B^n}_1$ ($n=1,2,3,4$) from the large-$z$ expansion of 
(\ref{R2_1}), and obtained the result suggesting supersymmetry invariance at the torus topology: 
\be
  \vev{\frac{1}{N}\tr\,B^n}_1=0  \qquad (n=1,2,3,4).  
\label{Bn_torus}  
\ee
In ref.~\cite{Kuroki:2009yg} 
we find strong evidences that the supersymmetry is preserved at the sphere topology 
in the asymmetric one-cut phase and in the two-cut phase both corresponding to \eqref{R2_0}. 
It is therefore interesting if the supersymmetry remains intact at the torus topology.

\section{Dependence on filling fractions of $\vev{\prod_{i}\frac{1}{N}\tr\,\phi^{n_i}}_C$}
\label{app:ff}
\setcounter{equation}{0}
We here discuss dependence on filling fractions $(\nu_+, \nu_-)$      
of the $K$-point correlation functions $\vev{\prod_{i=1}^K\frac{1}{N}\tr\,\phi^{n_i}}_C$ with $K=1,2,3$. 

Let us express as $Z_{(\nu_+,\nu_-)}$ contribution to the partition function (\ref{Z}) from 
the configurations where the first $\nu_+N$ eigenvalues of the matrix $\phi$ are around the right minimum 
and the remaining $\nu_-N$ around the left minimum:
\bea
& \lambda_i = \mu + \tilde{\lambda}_i & \mbox{for }i\in I\equiv\{1,\cdots, \nu_+N\},  \nn \\
& \lambda_j = -\mu + \tilde{\lambda}_j & \mbox{for }j\in J\equiv\{\nu_+N+1, \cdots, N\}
\label{nu+nu-_config}
\eea 
with $\tilde{\lambda}_i$ and $\tilde{\lambda}_j$ small fluctuations.  
Then, 
\be
Z  =   \sum_{\nu_+N=0}^N\frac{N!}{(\nu_+N)!(\nu_-N)!}\,Z_{(\nu_+,\nu_-)}  
\ee
is valid at least to all orders in perturbation with respect to the small fluctuations,   
and 
\bea
Z_{(\nu_+,\nu_-)} & = & \tilde{C}_N\int\left(\prod_{k=1}^Nd\tilde{\lambda}_k\right) 
\prod_{i\in I} W''(\mu+\tilde{\lambda}_i)\prod_{j\in J}W''(-\mu+\tilde{\lambda}_j) \nn \\
& & \hspace{14mm} \times 
\prod_{i>i', \,i,i'\in I} \left(W'(\mu+\tilde{\lambda}_i)- W'(\mu+\tilde{\lambda}_{i'})\right)^2 
\nn \\
& & \hspace{14mm}\times 
\prod_{j>j', \,j,j'\in J} \left(W'(-\mu+\tilde{\lambda}_j)- W'(-\mu+\tilde{\lambda}_{j'})\right)^2 \nn \\
& & \hspace{14mm}\times 
\prod_{i\in I, \,j\in J}\left(W'(-\mu+\tilde{\lambda}_j)- W'(\mu+\tilde{\lambda}_i)\right)^2 \nn \\
& & \hspace{14mm}\times 
\exp\left[-N\sum_{i\in I}\frac12W'(\mu+\tilde{\lambda}_i)^2-N\sum_{j\in J}\frac12W'(-\mu+\tilde{\lambda}_j)^2
\right]
\label{Z_nu+nu-}
\eea
with $W'(x)=x^2-\mu^2$, $W''(x)=2x$. $\tilde{C}_N$ is a numerical factor depending only on $N$ 
\cite{Kuroki:2010au}. 
The integrals in (\ref{Z_nu+nu-}) are calculated in perturbation with respect 
to the small fluctuations, namely in the $\frac{1}{N}$-expansion.  
 By flipping the sign of $\tilde{\lambda}_j$ ($j\in J$), it is easy to see 
\be
Z_{(\nu_+,\nu_-)} = (-1)^{\nu_-N} Z_{(1,0)}. 
\label{Z_nu+nu-_10}
\ee 

Let us write correlators evaluated under the configurations (\ref{nu+nu-_config}) as 
$\vev{\cdot}^{(\nu_+,\nu_-)}$. In the remaining of this appendix, we find a relation between 
$\vev{\prod_{i=1}^K\frac{1}{N}\tr\,\phi^{n_i}}_C^{(\nu_+,\nu_-)}$ and 
$\vev{\prod_{i=1}^K\frac{1}{N}\tr\,\phi^{n_i}}_C^{(1,0)}$ for $K=1,2,3$. 

\subsection{$K=1$}
\label{app:K=1}
Applying the same argument of flipping the sign 
to $\vev{\frac{1}{N}\tr\,\phi^{n}}^{(\nu_+,\nu_-)}$ leads to 
\be
 \vev{\frac{1}{N}\tr\,\phi^{n}}^{(\nu_+,\nu_-)}
 =
 \vev{\frac{1}{N}\sum_{i\in I}(\mu+\tilde{\lambda}_i)^n}^{(1,0)} 
 +(-1)^n \vev{\frac{1}{N}\sum_{j\in J}(\mu+\tilde{\lambda}_j)^n}^{(1,0)} . 
\ee
Since there is a permutation symmetry of $\tilde{\lambda}_1, \cdots, \tilde{\lambda}_N$ 
in $\vev{\cdot}^{(1,0)}$, 
we can write the r.h.s. as 
\be
(\nu_+ + (-1)^n \nu_-)\vev{(\mu+\tilde{\lambda}_1)^n}^{(1,0)} 
= (\nu_+ + (-1)^n \nu_-)\vev{\frac{1}{N}\tr \,\phi^n}^{(1,0)}. 
\ee
Thus 
\be
\vev{\frac{1}{N}\tr\,\phi^{n}}^{(\nu_+,\nu_-)} 
= (\nu_+ + (-1)^n \nu_-)\vev{\frac{1}{N}\tr \,\phi^n}^{(1,0)}.  
\label{nu+nu-_10_K=1}
\ee
Namely, $\vev{\frac{1}{N}\tr\,\phi^{n}}^{(\nu_+,\nu_-)} $ is independent of the filling fractions 
for $n$ even, and proportional to $(\nu_+-\nu_-)$ for 
$n$ odd. 
Note that the relation (\ref{nu+nu-_10_K=1}) holds at each order of the $\frac{1}{N}$-expansion. 

\subsection{$K=2$}
\label{app:K=2}
After the sign flip, we have 
\bea
& & \hspace{-7mm} \vev{\frac{1}{N}\tr\,\phi^{n_1}\,\frac{1}{N}\tr\,\phi^{n_2}}_C^{(\nu_+,\nu_-)} \nn \\
&  & =\vev{\frac{1}{N}\sum_{i_1\in I}(\mu+\tilde{\lambda}_{i_1})^{n_1}\,
\frac{1}{N}\sum_{i_2\in I}(\mu+\tilde{\lambda}_{i_2})^{n_2}}^{(1,0)}  
\label{1st_K=2}\\
 & & \hspace{4mm}+(-1)^{n_1+n_2}\vev{\frac{1}{N}\sum_{j_1\in J}(\mu+\tilde{\lambda}_{j_1})^{n_1}\,
\frac{1}{N}\sum_{j_2\in J}(\mu+\tilde{\lambda}_{j_2})^{n_2}}^{(1,0)}  
\label{2nd_K=2}\\
 & & \hspace{4mm}+ (-1)^{n_1} \vev{\frac{1}{N}\sum_{j_1\in J}(\mu+\tilde{\lambda}_{j_1})^{n_1}\,
\frac{1}{N}\sum_{i_2\in I}(\mu+\tilde{\lambda}_{i_2})^{n_2}}^{(1,0)}  
\label{3rd_K=2}\\
 & & \hspace{4mm}+(-1)^{n_2} \vev{\frac{1}{N}\sum_{i_1\in I}(\mu+\tilde{\lambda}_{i_1})^{n_1}\,
 \frac{1}{N}\sum_{j_2\in J}(\mu+\tilde{\lambda}_{j_2})^{n_2}}^{(1,0)}  
\label{4th_K=2}\\
 & & \hspace{4mm}-\left(\nu_++(-1)^{n_1}\nu_-\right)\left(\nu_++(-1)^{n_2}\nu_-\right) 
\vev{\frac{1}{N}\tr \,\phi^{n_1}}^{(1,0)}\vev{\frac{1}{N}\tr \,\phi^{n_2}}^{(1,0)}. 
\label{last_K=2}
\eea

Use of the permutation symmetry recasts the first term in the r.h.s. as 
\bea
(\ref{1st_K=2}) & = & \nu_+^2\vev{(\mu+\tilde{\lambda}_1)^{n_1}\, (\mu+\tilde{\lambda}_2)^{n_2}}^{(1,0)} \nn \\
& & + \frac{\nu_+}{2N}\vev{\left\{(\mu+\tilde{\lambda}_1)^{n_1}-(\mu+\tilde{\lambda}_2)^{n_1}\right\} 
 \left\{(\mu+\tilde{\lambda}_1)^{n_2}-(\mu+\tilde{\lambda}_2)^{n_2}\right\}}_C^{(1,0)}. \nn \\
\eea
Notice that the connected part is taken in the second term because 
the disconnected part vanishes due to the permutation symmetry. Since 
$(\nu_+,\nu_-)=(1,0)$ case in the above argument yields 
\bea
& & \hspace{-7mm} \vev{\frac{1}{N}\tr\,\phi^{n_1}\,\frac{1}{N}\tr\,\phi^{n_2}}^{(1,0)} = 
 \vev{(\mu+\tilde{\lambda}_1)^{n_1}\, (\mu+\tilde{\lambda}_2)^{n_2}}^{(1,0)} \nn \\
& &  + \frac{1}{2N}\vev{\left\{(\mu+\tilde{\lambda}_1)^{n_1}-(\mu+\tilde{\lambda}_2)^{n_1}\right\} 
 \left\{(\mu+\tilde{\lambda}_1)^{n_2}-(\mu+\tilde{\lambda}_2)^{n_2}\right\}}_C^{(1,0)},
\eea 
we obtain 
\bea
(\ref{1st_K=2}) & = & \nu_+^2 \vev{\frac{1}{N}\tr\,\phi^{n_1}\,\frac{1}{N}\tr\,\phi^{n_2}}^{(1,0)}\nn \\
& & + \frac{\nu_+\nu_-}{2N}\vev{\left\{(\mu+\tilde{\lambda}_1)^{n_1}-(\mu+\tilde{\lambda}_2)^{n_1}\right\} 
 \left\{(\mu+\tilde{\lambda}_1)^{n_2}-(\mu+\tilde{\lambda}_2)^{n_2}\right\}}_C^{(1,0)}.   \nn \\
 & & 
\label{1st_K=2_f} 
\eea

Repeating the same argument for (\ref{2nd_K=2}), (\ref{3rd_K=2}) and (\ref{4th_K=2}), and putting  
the results together with the last term (\ref{last_K=2}), we end up with 
\bea
& & \hspace{-7mm} \vev{\frac{1}{N}\tr\,\phi^{n_1}\,\frac{1}{N}\tr\,\phi^{n_2}}_C^{(\nu_+,\nu_-)} \nn \\
&  &=\left(\nu_++(-1)^{n_1}\nu_-\right)\left(\nu_++(-1)^{n_2}\nu_-\right) 
\vev{\frac{1}{N}\tr\,\phi^{n_1}\,\frac{1}{N}\tr\,\phi^{n_2}}^{(1,0)}_C\nn \\
& & \hspace{4mm} +\left(1-(-1)^{n_1}\right)\left(1-(-1)^{n_2}\right)\frac{\nu_+\nu_-}{2N} \nn \\
& & \hspace{7mm}\times\vev{\left\{(\mu+\tilde{\lambda}_1)^{n_1}-(\mu+\tilde{\lambda}_2)^{n_1}\right\} 
 \left\{(\mu+\tilde{\lambda}_1)^{n_2}-(\mu+\tilde{\lambda}_2)^{n_2}\right\}}_C^{(1,0)}.  
\label{nu+nu-_10_K=2} 
\eea 
This formula tells us that 
\be 
\vev{\frac{1}{N}\tr\,\phi^{n_1}\,\frac{1}{N}\tr\,\phi^{n_2}}_C^{(\nu_+,\nu_-)} 
= \vev{\frac{1}{N}\tr\,\phi^{n_1}\,\frac{1}{N}\tr\,\phi^{n_2}}_C^{(1,0)} 
\label{nu+nu-_10_K=2_ee}
\ee
for $n_1$ and $n_2$ even, and 
\be
\vev{\frac{1}{N}\tr\,\phi^{n_1}\,\frac{1}{N}\tr\,\phi^{n_2}}_C^{(\nu_+,\nu_-)} 
= (\nu_+-\nu_-)\vev{\frac{1}{N}\tr\,\phi^{n_1}\,\frac{1}{N}\tr\,\phi^{n_2}}_C^{(1,0)} 
\label{nu+nu-_10_K=2_eo}
\ee
for one of $n_1$ and $n_2$ odd. These are valid to all orders in the $\frac{1}{N}$-expansion. 
For $n_1$ and $n_2$ odd, 
the second term in the r.h.s. of (\ref{nu+nu-_10_K=2}) appears to be nonvanishing. However, it is 
at most of $\cO(N^{-3})$ and negligible compared to the leading contribution of $\cO(N^{-2})$. 
Thus we have shown that   
\be
 \vev{\frac{1}{N}\tr\,\phi^{n_1}\,\frac{1}{N}\tr\,\phi^{n_2}}_{C, 0}^{(\nu_+,\nu_-)} 
= (\nu_+-\nu_-)^2\vev{\frac{1}{N}\tr\,\phi^{n_1}\,\frac{1}{N}\tr\,\phi^{n_2}}_{C,0}^{(1,0)} 
\label{nu+nu-_10_K=2_oo_planar}
\ee
for $n_1$ and $n_2$ odd.

\subsection{$K=3$}
\label{app:K=3}
The argument parallel to the previous subsection leads to the result 
\bea
 & & \hspace{-7mm}\vev{\prod_{i=1}^3\frac{1}{N}\tr\,\phi^{n_i}}_C^{(\nu_+,\nu_-)} 
= \left\{\prod_{i=1}^3\left(\nu_++(-1)^{n_i}\nu_-\right)\right\}
\vev{\prod_{i=1}^3\frac{1}{N}\tr\,\phi^{n_i}}_C^{(1,0)} \nn \\
 & & + \frac{\nu_+\nu_-}{2N}\left[\left(1-(-1)^{n_1}\right)\left(1-(-1)^{n_2}\right)
\left(\nu_++(-1)^{n_3}\nu_-\right) \frac{}{}\right. \nn \\
& & \hspace{21mm} \times \vev{\left\{(\mu+\tilde{\lambda}_1)^{n_1}-(\mu+\tilde{\lambda}_2)^{n_1}\right\} 
 \left\{(\mu+\tilde{\lambda}_1)^{n_2}-(\mu+\tilde{\lambda}_2)^{n_2}\right\}
(\mu+\tilde{\lambda}_3)^{n_3}}_C^{(1,0)} \nn \\
& & \left.\frac{}{}\hspace{14mm} + (\mbox{two terms from cyclic permutations of }(n_1, n_2, n_3))\right] \nn \\
& & + \frac{2\nu_+\nu_-}{3N^2}f(n_1,n_2,n_3) \left[\left\bra
\left\{(\mu+\tilde{\lambda}_1)^{n_1}-(\mu+\tilde{\lambda}_2)^{n_1}\right\} 
 \left\{(\mu+\tilde{\lambda}_1)^{n_2}-(\mu+\tilde{\lambda}_2)^{n_2}\right\} \right.\right. \nn \\
& & \hspace{41mm}\left. \times \left\{(\mu+\tilde{\lambda}_1)^{n_3}+(\mu+\tilde{\lambda}_2)^{n_3} 
-2(\mu+\tilde{\lambda}_3)^{n_3}\right\}\right\ket_C^{(1,0)} \nn \\
& & \left.\frac{}{}\hspace{35mm} + (\mbox{two terms from cyclic permutations of }(n_1, n_2, n_3))\right],  
\label{nu+nu-_10_K=3}
\eea
where 
\bea
f(n_1,n_2,n_3) = \begin{cases} 0 & \mbox{(two or all of $n_1, \,n_2, \,n_3$ are even)}, \\
                               1 & \mbox{(one of $n_1, \,n_2, \,n_3$ is even)}, \\
                               \nu_+-\nu_- & \mbox{(all of $n_1, \,n_2, \,n_3$ are odd)}.
\end{cases} 
\eea

{}From (\ref{nu+nu-_10_K=3}), we see that 
\be
\vev{\prod_{i=1}^3\frac{1}{N}\tr\,\phi^{n_i}}_C^{(\nu_+,\nu_-)} 
= \vev{\prod_{i=1}^3\frac{1}{N}\tr\,\phi^{n_i}}_C^{(1,0)}
\label{nu+nu-_10_K=3_eee}
\ee
for $n_1$, $n_2$ and $n_3$ all even, and 
\be 
\vev{\prod_{i=1}^3\frac{1}{N}\tr\,\phi^{n_i}}_C^{(\nu_+,\nu_-)} 
= (\nu_+-\nu_-)\vev{\prod_{i=1}^3\frac{1}{N}\tr\,\phi^{n_i}}_C^{(1,0)}
\label{nu+nu-_10_K=3_eeo}
\ee
for one of $n_1$, $n_2$ and $n_3$ odd. These hold to all orders in the $\frac{1}{N}$-expansion. 
For the other cases, although the second and third terms in the r.h.s. of (\ref{nu+nu-_10_K=3}) 
appear to remain nonzero, 
they are at most of $\cO(N^{-5})$. So, by looking at the leading order of $\cO(N^{-4})$, 
\be
 \vev{\prod_{i=1}^3\frac{1}{N}\tr\,\phi^{n_i}}_{C,0}^{(\nu_+,\nu_-)} 
= (\nu_+-\nu_-)^\sharp\vev{\prod_{i=1}^3\frac{1}{N}\tr\,\phi^{n_i}}_{C,0}^{(1,0)}
\label{nu+nu-_10_K=3_ooo_planar}
\ee
is shown. Here, $\sharp$ is the number of odd integers in $\{n_1, n_2, n_3\}$. 

{}From the results obtained in this section, it is expected that 
$\vev{\prod_{i=1}^K\frac{1}{N}\tr\,\phi^{n_i}}_{C,0}$ for general $K$ 
is proportional to $(\nu_+-\nu_-)^\sharp$ with 
$\sharp$ the number of odd-power operators of $\phi$.

\section{Useful formulas}
\label{app:formulas}
\setcounter{equation}{0}
The semi-circle eigenvalue distribution $\sqrt{1-y^2}$ ($y\in[-1,1]$) 
appearing in the Gaussian one-matrix model $S=N\tr\,(2\phi^2)$ satisfies the saddle point equation 
\be
\int^1_{-1}dy\,\sqrt{1-y^2}\,{\rm P}\frac{1}{x-y} = \pi x 
\qquad \mbox{for }x\in[-1,1], 
\label{app:GMM0}
\ee
from which we obtain 
\bea
& & \int^1_{-1}dy\,\sqrt{1-y^2}\,{\rm P}\frac{1}{x-y}\,y \nn \\
& & = x\int^1_{-1}dy\,\sqrt{1-y^2}\,{\rm P}\frac{1}{x-y} -\int^1_{-1}dy\,\sqrt{1-y^2} = \pi\left(x^2-\frac12\right). 
\label{app:GMM1} 
\eea

Next, let us compute 
\be
\int^1_{-1}dy\,\sqrt{1-y^2}\,{\rm P}\frac{1}{x-y}\,{\rm P}\frac{1}{u-y}, \qquad x, u\in[-1,1]. 
\label{app:formula1}
\ee 
We express the principal values as 
\bea
{\rm P}\frac{1}{x-y} & = & \frac12\left(\frac{1}{x-y+i\epsilon} + \frac{1}{x-y-i\epsilon}\right), \nn \\
{\rm P}\frac{1}{u-y} & = & \frac12\left(\frac{1}{u-y+i\epsilon'} + \frac{1}{u-y-i\epsilon'}\right)  
\eea
with $\epsilon$ and $\epsilon'$ infinitesimal, and rewrite  
${\rm P}\frac{1}{x-y}\,{\rm P}\frac{1}{u-y}$ as 
\bea
{\rm P}\frac{1}{x-y}\,{\rm P}\frac{1}{u-y} & = &  \frac14\left[
 \frac{1}{x-y+i\epsilon}\frac{1}{u-y+i\epsilon'} + \frac{1}{x-y-i\epsilon}\frac{1}{u-y-i\epsilon'}
\right. \nn \\
& & \hspace{3mm} 
\left. + \frac{1}{x-y+i\epsilon}\frac{1}{u-y-i\epsilon'} + \frac{1}{x-y-i\epsilon}\frac{1}{u-y+i\epsilon'}
\right]. 
\label{app:PP_henkei}
\eea
We recast the first term as 
\bea
& & \frac{1}{x-y+i\epsilon}\frac{1}{u-y+i\epsilon'}
= \left(\frac{1}{x-y+i\epsilon} - \frac{1}{u-y+i\epsilon'}\right)\frac{1}{u-x+i(\epsilon'-\epsilon)} \nn \\
& & = \left({\rm P}\frac{1}{x-y}-i\pi\delta(x-y) -{\rm P}\frac{1}{u-y}+i\pi\delta(u-y)\right) 
\frac{1}{u-x+i(\epsilon'-\epsilon)},   
\eea
and carry out the integral in the l.h.s. of (\ref{app:formula1}) by using (\ref{app:GMM0}). 
After doing in the same manner for the other terms in (\ref{app:PP_henkei}), we arrive at 
\bea
(\ref{app:formula1}) & = & -\pi +i\frac{\pi}{4}\sqrt{1-x^2}
\left(-\frac{1}{u-x+i(\epsilon'-\epsilon)} + \frac{1}{u-x+i(-\epsilon'+\epsilon)} \right. \nn \\
& & \left.\hspace{30mm}
-\frac{1}{u-x+i(-\epsilon'-\epsilon)} + \frac{1}{u-x+i(\epsilon'+\epsilon)} \right) \nn \\
& & +i\frac{\pi}{4}\sqrt{1-u^2}
\left(\frac{1}{u-x+i(\epsilon'-\epsilon)} - \frac{1}{u-x+i(-\epsilon'+\epsilon)} \right. \nn \\
& & \left. \hspace{24mm} -\frac{1}{u-x+i(-\epsilon'-\epsilon)} + \frac{1}{u-x+i(\epsilon'+\epsilon)} \right). 
\label{app:formula2}
\eea
Here, since (\ref{app:formula2}) becomes equal to $-\pi + \pi^2\sqrt{1-u^2}\,\delta(u-x)$ 
in the limit $\epsilon, \epsilon' \to 0$ with either $\epsilon > \epsilon' >0$ or $\epsilon'>\epsilon>0$, 
we conclude that 
\be
\int^1_{-1}dy\,\sqrt{1-y^2}\,{\rm P}\frac{1}{x-y}\,{\rm P}\frac{1}{u-y}
= -\pi + \pi^2\sqrt{1-u^2}\,\delta(u-x) 
\label{app:formula}
\ee
for $x, u\in [-1,1]$. 

Finally, we show 
\be
\int^1_{-1}dy\,\frac{1}{\sqrt{1-y^2}}\,{\rm P}\frac{1}{x-y}=0
\qquad \mbox{for} \quad x\in[-1,1]. 
\label{app:formula3}
\ee
The l.h.s. is written as 
\bea
(\mbox{l.h.s.}) & = & \frac12\frac{1}{1-x}\left[\int^1_{-1}dy\sqrt{1-y^2}\left({\rm P}\frac{1}{x-y}-\frac{1}{1-y}\right)\right] \nn \\
 & & 
+\frac12\frac{1}{1+x}\left[\int^1_{-1}dy\sqrt{1-y^2}\left({\rm P}\frac{1}{x-y}-\frac{1}{-1-y}\right)\right]. 
\eea
Applying (\ref{app:GMM0}) leads to 
\be
(\mbox{l.h.s.}) = \frac12\frac{1}{1-x}(\pi x-\pi)+ \frac12\frac{1}{1+x}(\pi x+\pi) = 0. 
\ee


\end{document}